\documentclass[a4paper,11pt]{article}
\usepackage{jheppub}

\usepackage{chessfss}
\pdfoutput=1
\pdfpageattr {/Group << /S /Transparency /I true /CS /DeviceRGB>>}
\usepackage{hyperref}
\hypersetup{colorlinks=true,linkcolor=blue,citecolor=blue}
\usepackage{amsmath,amssymb,amsfonts}
\usepackage{bm}
\usepackage{float}
\usepackage{subfig}
\usepackage{subfig}
\usepackage{scalefnt}
\usepackage{wrapfig}
\usepackage{fancybox}
\usepackage{caption}
\usepackage{tikz}
\usepackage{multirow}

\usetikzlibrary{arrows}
\tikzset{>=angle 60}
\tikzstyle{W}=[draw,circle,scale=.6]
\tikzstyle{B}=[draw,circle,fill=black,scale=.6]
\tikzstyle{H}=[draw,circle,fill=gray,scale=.6]
\tikzstyle{every picture}=[scale=.6,baseline=(current bounding box.south)]

\def\beq#1\eeq{\begin{align}#1\end{align}}

\newcommand{\be}{\begin{eqnarray}}
\newcommand{\ee}{\end{eqnarray}}
\newcommand{\bea}{\begin{eqnarray}}
\newcommand{\eea}{\end{eqnarray}}

\newcommand{\bn}{\begin{enumerate}}
\newcommand{\en}{\end{enumerate}}

\newcommand{\ada}[4]{\ensuremath{\frac{a^{(#1)}_{#2}}{a^{(#3)}_{#4}}}}
\newcommand{\xb}[3]{\ensuremath{x_{#1}^{(#2,#3)}}}
\newcommand{\yb}[3]{\ensuremath{y_{#1}^{(#2,#3)}}}

\def\half{\frac{1}{2}}

\title{ADE String Chains and Mirror Symmetry}

\begin{document}
\author[\ast]{Babak Haghighat,}
\author[\ast, \dag]{Wenbin Yan\footnote{Primary affiliation: Yau Mathematical Sciences Center, Tsinghua University, Beijing, China},}
\author[\ddagger]{Shing-Tung Yau}
\affiliation[\ast]{Yau Mathematical Sciences Center, Tsinghua University, Beijing, 100084, China}
\affiliation[\dag]{Center of Mathematical Sciences and Applications, Harvard University, Cambridge, 02138, USA}
\affiliation[\ddagger]{Department of Mathematics, Harvard University, Cambridge, 02138, USA}

\abstract{6d superconformal field theories (SCFTs) are the SCFTs in the highest possible dimension. They can be geometrically engineered in F-theory by compactifying on non-compact elliptic Calabi-Yau manifolds. In this paper we focus on the class of SCFTs whose base geometry is determined by $-2$ curves intersecting according to ADE Dynkin diagrams and derive the corresponding mirror Calabi-Yau manifold. The mirror geometry is uniquely determined in terms of the mirror curve which has also an interpretation in terms of the Seiberg-Witten curve of the four-dimensional theory arising from torus compactification. Adding the affine node of the ADE quiver to the base geometry, we connect to recent results on SYZ mirror symmetry for the $A$ case and provide a physical interpretation in terms of little string theory. Our results, however, go beyond this case as our construction naturally covers the $D$ and $E$ cases as well.}
\maketitle

\section{Introduction}

By now there is compelling evidence that 6d $\mathcal{N}=(1,0)$ superconfornal field theories (SCFTs) are classified in terms of non-compact elliptic Calabi-Yau manifolds \cite{Heckman:2013pva,DelZotto:2014hpa,Heckman:2015bfa,Bhardwaj:2015oru,Heckman:2015ola}. It is therefore natural to initiate a classification program for the corresponding mirror Calabi-Yau manifolds. In physics language, this means classifying all Seiberg-Witten geometries which arise from two-torus compactifications of 6d SCFTs. There has been a first but incomplete attempt to do this \cite{DelZotto:2015rca} (for a yet earlier relevant work see \cite{Hollowood:2003cv}), where the authors largely focus on the class of conformal matter theories studied in \cite{DelZotto:2014hpa}.

In the present paper we focus on the class of SCFTs arising from compactification on Calabi-Yau manifolds with a base geometry of $-2$ curves intersecting according to $A$, $D$ or $E$ type Dynkin diagrams and an $A$ type elliptic fiber geometry. We derive the corresponding Seiberg-Witten geometries for non-affine base geometries and connect to SYZ mirror symmetry in the case of affine base geometries. Partition functions for such theories had been previously obtained in \cite{Gadde:2015tra} by computing elliptic genera of strings which arise on the tensor branch. Such elliptic genera are computed through a localization computation in a 2d quiver gauge theory and the results can be fully expressed in terms of summations over configurations of Young-diagrams. Building on earlier work \cite{Haghighat:2016jjf}, this allows us to use the thermodynamic limit technique of \cite{Nekrasov:2003rj} to compute the emerging geometry in the density limit of the Young diagrams. Our result provides a generalization of the work \cite{Nekrasov:2012xe} for four-dimensional quiver gauge theories to the six-dimensional setting which introduces novel features such as invariance under fiber-base duality and connections to SYZ mirror symmetry. Our work captures in the affine case the Seiberg-Witten geometries for little string theories which are UV complete non-local 6d theories decoupled from gravity. Coupling constants and Coulomb branch parameters of these theories then provide a parametrization for the family of mirror curves. In other words, the dimension of the moduli space of the mirror curve is given by the total number of parameters in the corresponding little string theory.

In a sense, this work brings together and connects three different papers, namely the computation of elliptic genera for ADE string chains of \cite{Gadde:2015tra}, the Seiberg-Witten geometries obtained from the thermodynamic limit of four-dimensional quiver gauge theories \cite{Nekrasov:2012xe}, and lastly recent mathematical results on SYZ mirrors of toric Calabi-Yau manifolds of infinite type \cite{Kanazawa:2016tnt}. The latter results correspond to taking the base geometry of our Calabi-Yau to be of affine $A$ type and indeed we recover the results of \cite{Kanazawa:2016tnt} from our point of view which we demonstrate in section \ref{sec:mirrorcurvederivagion} for the mass-less case. In fact, our results go beyond those of \cite{Kanazawa:2016tnt} since we also provide expressions of mirror curves for cases where the base geometry is of affine $D$ and $E$ type. These geometries have no toric realization and thus the expressions we provide should have interesting interpretations from the point of view of SYZ mirror symmetry in the non-toric setup.

Let us now come to the organization of the paper. In section \ref{sec:6dSCFT} we review the construction of the particular 6d SCFTs we are interested in, both in terms of brane configurations as well as in the framework of geometric engineering. This includes deriving the quiver gauge theory description in four and five dimensions from a Lagrangian point of view. Moreover, we interpret the instanton contributions to the corresponding BPS partition functions on $T^2 \times \mathbb{R}^4$ as self-dual strings wrapping the $T^2$. The worldsheet anomaly polynomials of these strings are then mapped to modular anomalies of their elliptic genera. Proceeding to section \ref{sec:ADE}, we derive a novel representation for the elliptic genera obtained in \cite{Gadde:2015tra}, which in turn allows us to take the thermodynamic limit of the full partition function. The last subsections of section \ref{sec:ADE} then deal with the resulting Seiberg-Witten geometry. Finally, in section \ref{sec:mirrorsymmetry} we connect to the results of \cite{Kanazawa:2016tnt} by adding the affine node to the base geometry. This construction is then interpreted in the context of little string theories before proceeding to a more explicit description of the resulting mirror curves.

\section{6d SCFTs from ADE singularities}
\label{sec:6dSCFT}

In this section, we review the structure of 6d $\mathcal{N}=(1,0)$ SCFTs arising from ADE configurations of $-2$ curves and their compactifications to four and five dimensions on $S^1$ and $T^2$.

\subsection{Geometric Engineering}

The class of theories we want to focus on in this section is the one studied in Section 5 of \cite{Gadde:2015tra}. This class of 6d SCFTs can be constructed by compactifying F-theory on elliptic Calabi-Yau threefolds with the following geometric properties of base and fiber. The base $B$ is a non-compact, complex two-dimensional space which is obtained by blowing up an ADE singularity. As such, it has 2-cycles $C^i$ which are $\mathbb{P}^1$'s with negative intersection matrix $\eta^{ij} = -C^i \cdot C^j$ being equal to the Cartan matrix of a simply laced gauge group of ADE type. Furthermore, above each $C^i$ we let the elliptic fiber degenerate according to an $I_{N_i}$ Kodaira singularity. In fact, $N_i$ will vary for each $\mathbb{P}^1$ and is proportional to the Dynkin label of the corresponding node in the ADE Dynkin diagram as will be explained in more detail in section \ref{sec:ADE}.

The resulting theory in 6d admits $h^{1,1}(B)$ $\mathcal{N}=(1,0)$ tensor multiplets and each $C^i$ further supports a gauge group $SU(N_i)$, and there are bifundamental hypermultiplets between curves which are intersecting. The resulting quiver gauge theory can be equivalently obtained from Type IIB string theory with $N$ D5 branes probing an asymptotically locally flat (ALF) singularity of ADE type as follows from the Douglas-Moore construction \cite{Douglas:1996sw}. In fact, the Type IIB setup can be shown to be dual to the F-theory compactification presented above. Following the notation of \cite{Ohmori:2015pia}, we will henceforth denote these theories by $\mathcal{T}^{\textrm{6d}}_{\mathfrak{g}}\{\textrm{SU}(N)\}$ where $\mathfrak{g}$ is the corresponding Lie algebra of A, D, or E type.

\subsection{$S^1$ compactification to five dimensions}

In the following we want to construct the tensor branch effective action upon compactification of our 6d theory on a circle. The bosonic components of the tensor multiplets are denoted by $(\varphi_i,B_i)$, where $\varphi_i$ are real scalars and $B_i$ are 2-forms whose field strengths are self-dual. The volume of the $(-2)$-curve labeled by $i$ is proportional to $\varphi^i = \eta^{ij} \varphi_j$, and the gauge field strength at the node $i$ due to seven-branes wrapping $C^i$ is denoted by $F_i$. Following \cite{Ohmori:2015pia}, a part of the formal bosonic effective action in six dimensions is given by
\begin{equation}
	2\pi \int \eta^{ij} \left(-\half d\varphi_i \wedge \star d \varphi_j - \half d B_i \wedge \star d B_j + \varphi_i(\frac{1}{4} \textrm{Tr} F_j \wedge \star F_j) + B_i(\frac{1}{4}\textrm{Tr} F_j \wedge F_j)\right).
\end{equation}
Note that the part containing the 2-form $B_i$ is required by Green-Schwarz anomaly cancellation, and the part containing $\varphi_i$ is related to the $B_i$ part by supersymmetry \cite{Green:1984bx,Sadov:1996zm,Blum:1997mm,Riccioni:1998th,Grassi:2011hq,Ohmori:2015pia}.

Upon dimensional reduction to 5d, we arrive at
\begin{equation}
	\int \eta^{ij} \left(-\frac{1}{2R}(d \phi_i \wedge \star d \phi_j + dA_i \wedge \star d A_j) + 2\pi \phi_i (\frac{1}{4} F_j \wedge \star F_j) + 2\pi A_i (\frac{1}{4} \textrm{Tr} F_j \wedge F_j)\right),
\end{equation}
where $\phi_i$ and $A_i$ are defined as follows
\begin{equation}
	\phi_i = 2\pi R \varphi^i, \quad A^i_{\mu} = 2\pi R B^i_{\mu 5}.
\end{equation}
We see that the gauge couplings $g_i^2$ at each node $i$ of the resulting quiver gauge theory are determined by the vevs of the scalars of the 6d tensor multiplets, namely
\begin{equation}
	\frac{8 \pi^2}{g_i^2} = \phi_i.
\end{equation}
This identification will be crucial later on when we compute the Nekrasov partition functions of the resulting lower dimensional gauge theory.

\subsection{$T^2$ compactification to four dimensions}

When we compactify further to four dimensions we obtain a conformal quiver gauge theory with gauge couplings
\begin{equation}
	\rho_i = \textrm{Vol}(T^2)(i \varphi^i + B^i_{45}).
\end{equation}
The running of the gauge coupling $\rho_i$ is one-loop exact for $\mathcal{N}=2$ supersymmetric theories and is described by the following contribution of matter and gauge multiplets \cite{Nekrasov:2012xe,Novikov:1983uc}
\begin{equation}
	\beta_i = 2\pi i \frac{d\rho_i}{d\log \Lambda} = - 2 N_i + \sum_{e: t(e)=i} N_{s(e)} + \sum_{e: s(e)=i} N_{t(e)},
\end{equation}
where we are following the notation of \cite{Nekrasov:2012xe}. That is, $\Lambda$ is the energy scale and the sum is over oriented edges $e$ of the quiver which end or start at the node $i$ and $s(e)$/$t(e)$ are its source/target nodes. The expression above can be simplified by introducing the oriented \textit{adjacency} matrix $M_{ij}$ which counts oriented edges between nodes $i$ and $j$:
\begin{equation}
	\beta_i = \sum_j  (-2\delta_{ij} + M_{ij})N_j = -\sum_j C_{ij} N_j,
\end{equation}
where $C_{ij}$ is the Cartan matrix associated to the quiver.
It follows naturally from the Type IIB Douglas-Moore construction presented above that $N_i = N d_i$ for all $i$, where $d_i$ are the Dynkin indices of the node $i$. Hence we have
\begin{equation}
\label{eq:anoCond}
	\beta_i = - N \sum_{j} C_{ij} d_j = 0,
\end{equation}
and thus we see that all couplings are conformal.

\subsection{Instanton strings}

Our goal in this paper will be to study instanton partition functions of the above described compactified quiver gauge theories in the thermodynamic limit. In order to proceed we will need to identify BPS instanton contributions to the four-dimensional partition function. These are given by D3-branes wrapping four-cycles $C^i \times T^2$. From the point of view of the six-dimensional SCFT on its tensor branch these are strings with tension $\varphi^i$. Upon compactification on $T^2$ these strings become instantons of $SU(N_i)$ and contribute as such to the BPS partition function of the resulting four-dimensional theory.

The world-volume theory of the strings is studied in \cite{Gadde:2015tra} and is described by a quiver gauge theory. The elliptic genus of this quiver gauge theory can be refined with respect to R- and flavor-symmetries. In particular, it is dependent on $\epsilon_1$ and $\epsilon_2$ which are the chemical potentials of rotations of $\mathbb{R}^4$ inside the worldvolume of the 6d SCFT. Furthermore, there will be dependencies on the $SU(N_i)$ fugacities denoted by $\mu^{(i)}_l$. Denoting the elliptic genus by
\begin{equation}
	Z_{\vec{k}}(\tau,\epsilon_1,\epsilon_2,\mu^{(i)}_l),
\end{equation}
where $\tau$ is the complex structure of $T^2$, one finds that the $T^2 \times \mathbb{R}^4$ partition function of the 6d SCFT on its tensor branch can be written as \cite{Haghighat:2013gba}
\begin{equation}
	Z_{T^2 \times \mathbb{R}^4}(\tau, \rho_i, \epsilon_1,\epsilon_2,\mu^{(i)}_l) = \sum_{\vec{k}} e^{2\pi i \sum_{i}k_i \rho_i} Z_{\vec{k}}(\tau,\epsilon_1,\epsilon_2,\mu^{(i)}_l),
\end{equation}
where $k_i$ denotes the number of D3-branes wrapping the cycle $C^i$. As argued in \cite{Witten:2009at} (see also \cite{DelZotto:2015isa}), six-dimensional theories with self-dual two-forms do not admit a scalar partition function but rather the partition function can be interpreted as an element of a vector space. This is related to the fact that the center $\mathcal{Z}$ of the Lie group whose Cartan matrix is given by the intersection pairing, is non-trivial for ADE type intersections except for $E_8$. In our case, due to the choice of background geometry, this vector space collapses to a one-dimensional Hilbert space and thus the partition function behaves as a scalar. More specifically, our background geometry $M_6 = M_4 \times S^1 \times \widetilde{S}^1$ can be viewed in more than one way as the product of a circle and a five-manifold. We can write $H^3(M_6,\mathcal{Z}) = A \oplus B$, where 
\begin{equation}
	A = H^2(M_4,\mathcal{Z}) \otimes H^1(S^1,\mathbb{Z}), \quad B = H^2(M_4,\mathcal{Z}) \otimes H^1(\widetilde{S}^1,\mathbb{Z}).
\end{equation}
As argued in \cite{Witten:2009at}, the relation between these two bases is given by a Fourier transform and in the present case we have the following relation between the partition vectors
\begin{equation}
	\widetilde{Z}_{{M_6},b} = C \sum_{a \in H^2(M_4,\mathbb{Z})} \exp(2\pi i (a,b))Z_{{M_6},a}.
\end{equation}
In the above, $C$ is a constant and $\exp(2\pi i (a,b))$ denotes the perfect pairing $H^2(M_4,\mathcal{Z}) \times H^2(M_4,\mathcal{Z}) \rightarrow U(1)$. Now in our specific case the four-manifold $M_4$ is given by the $\Omega$-background $\mathbb{R}^4_{\epsilon_1,\epsilon_2}$. This choice uniquely specifies one distinguished class in $H^2(M_4,\mathcal{Z})$ and collapses the sum to a single entry.

\subsubsection*{Anomalies}
The worldvolume theory of the strings suffers from anomalies with respect to global symmetries $SU(2)_L \times SU(2)_R \times SU(2)_I \times \prod_a G_a$. Here $SU(2)_L \times SU(2)_R \equiv SO(4)_N$ comes from rotating the normal directions to the string, while $SU(2)_I \times \prod_a G_a$ comes from the R, gauge and global symmetries of the bulk 6d theory. As shown in \cite{Shimizu:2016lbw} (see also \cite{DelZotto:2016pvm,Apruzzi:2016nfr}) the corresponding anomaly polynomial can be computed through the inflow formalism and the result is as follows
\begin{equation}
	I_4 = \frac{\eta^{ij}k_i k_j}{2}\left(c_2(L) - c_2(R)\right) + k_i \left(\frac{1}{4}\eta^{ij} \textrm{Tr} F_j^2 - \frac{2-\eta^{ii}}{4}(p_1(T)-2 c_2(L) - 2 c_2(R)) + h^{\vee}_{G_i} c_2(I) \right).
\end{equation}
There is yet another anomaly corresponding to modular transformations of $Z_{\vec{k}}$ and was already discussed in \cite{Haghighat:2013gba}:
\begin{equation}
	Z_{\vec{k}}(-1/\tau,\epsilon_i/\tau,\mu^{(i)}_l/\tau) = e^{\frac{2\pi i}{\tau}f(\epsilon_i,\mu^{(i)}_l)} Z_{\vec{k}}(\tau,\epsilon_i,\mu_i).
\end{equation}
As remarked in \cite{DelZotto:2016pvm} these two anomalies are related by performing the substitutions
\begin{equation}
	c_2(R) \rightarrow -\epsilon_+^2 \quad c_2(L)
	\rightarrow -\epsilon_-^2 \quad c_2(I) \rightarrow -\epsilon_+^2 \quad \frac{1}{4} \textrm{Tr}F_j^2 \rightarrow \mu^{(j)} \cdot \mu^{(j)},
\end{equation}
where $\mu^{(i)}$ denote the fugacities associated to the global symmetry group $G$ and $\epsilon_{\pm}=\frac{\epsilon_1 \pm \epsilon_2}{2}$. In our case, $G=SU(N_i)$ and $\eta^{ii}=2$. Using $h^{\vee}_{SU(N_i)}=N_i$ we then get
\begin{equation} \label{eq:anomaly}
	f(\epsilon_i,\mu^{(i)}_l) = \frac{\eta^{ij} k_i k_j}{2} \epsilon_1 \epsilon_2 + k_i\left(\eta^{ij} \mu^{(j)} \cdot \mu^{(j)} - N_i \epsilon_+^2\right).
\end{equation}

\section{Partition functions and their thermodynamic limit}
\label{sec:ADE}

In this section we  compute  partition functions on $T^2 \times \mathbb{R}^4$ for the general ADE case and express these in terms of elliptic genera of instanton strings.
In order to be able to treat the general ADE case we first need to clarify some definitions. To begin with, we note that the general quiver governing the theory of the self-dual strings \cite{Gadde:2015tra} consists of an outer and an inner quiver. The outer quiver, being an affine quiver, captures the gauge theory in the six-dimensional bulk and consists of flavor nodes from the viewpoint of the theory on the strings, while the inner quiver which is a standard one without affine nodes captures the gauge groups on the string world-sheet. In the following we will summarize our definitions and conventions.

\begin{figure}[h!]
  \centering
	\includegraphics[width=0.5\textwidth]{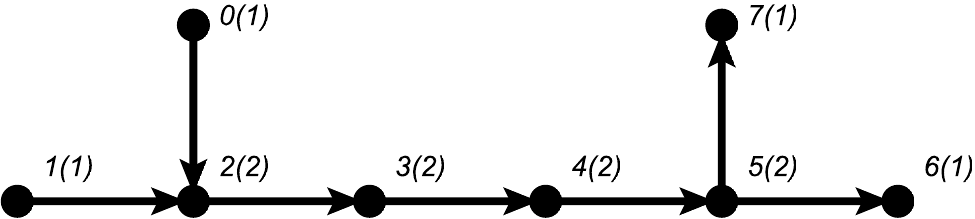}
  \caption{Oriented $\widehat{D}_7$ quiver. The numbers in the parentheses represent the Coxeter labels $d_i$.}
  \label{fig:AffineDquiver}
\end{figure}

\begin{figure}[h!]
  \centering
	\includegraphics[width=0.5\textwidth]{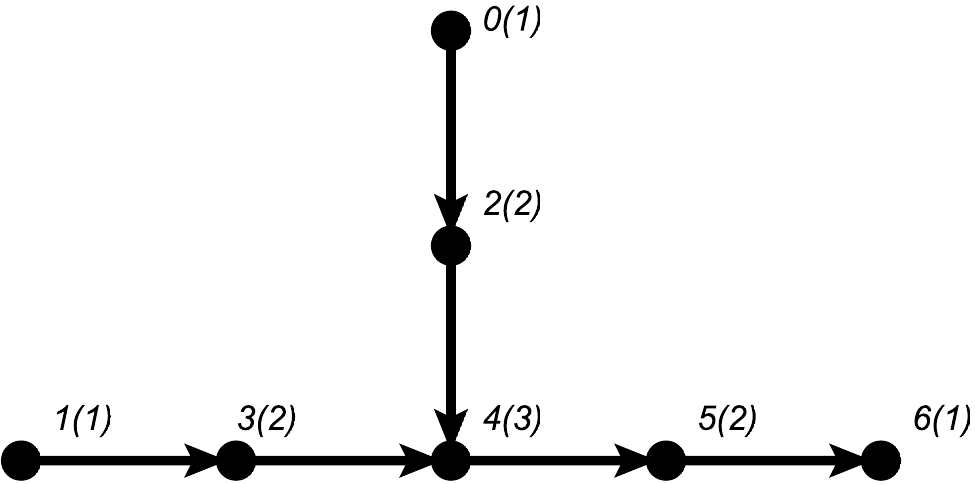}
  \caption{Oriented $\widehat{E}_6$ quiver. The numbers in the parentheses represent the Coxeter labels $d_i$.}
  \label{fig:AffineE6quiver}
\end{figure}

\begin{figure}[h!]
  \centering
	\includegraphics[width=0.7\textwidth]{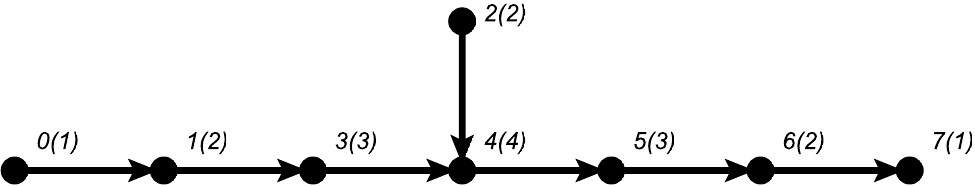}
  \caption{Oriented $\widehat{E}_7$ quiver. The numbers in the parentheses represent the Coxeter labels $d_i$.}
  \label{fig:AffineE7quiver}
\end{figure}

\begin{figure}[h!]
  \centering
	\includegraphics[width=0.8\textwidth]{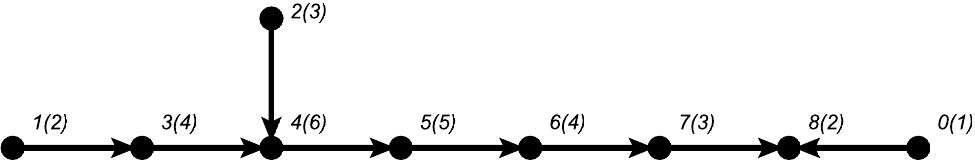}
  \caption{Oriented $\widehat{E}_8$ quiver. The numbers in the parentheses represent the Coxeter labels $d_i$.}
  \label{fig:AffineE8quiver}
\end{figure}

A node $i$ of the affine quiver with Coxeter label $d_i$ is associated with a gauge group $SU(d_i N)$, and each edge between nodes $i$ and $j$ is associated with bifundamental matter under $SU(d_i N)\times SU(d_j N)$. We denote by $a_l^{(i)} = e^{\mu_l^{(i)}}, l = 1, \ldots, d_i N$ the exponentiated fugacities corresponding to the flavor symmetry group $SU(d_i N)$ with the constraint $\prod a_l^{(i)} =1$. Furthermore, we will need for our expressions the adjacency matrix $M_{ij}$ of the affine quiver with \textit{some} orientation. Below we show explicit realizations of the matrix $M_{ij}$ for $A$, $D$ and $E$ type quivers. The $A$ case is different than the $D$ and $E$ cases as we split the extra node in the affine $A$ type quiver into two extra nodes at each end of the ordinary quiver with no connection between them. The adjacency matrix becomes
\begin{equation}
	M^{A_{r}} = \left(\begin{array}{cccccc}
		0 & 1 & 0 & \cdots & 0 & 0 \\
		0 & 0 & 1 & 0 & \cdots & 0 \\
		\vdots & ~ & ~ & \ddots & ~ & ~ \\
		0 & \cdots & 0 & 0 & 0 & 1 \\
		0 & \cdots & 0 & 0 & 0 & 0
	\end{array}\right)
\end{equation}
The indices $i,j$ run from $0$ to $r+1$ and all Coxeter labels $d_i$ are equal to $1$. In the case of a $\widehat{D}_r$-quiver we can read off the adjacency matrix from the following oriented quiver diagram depicted in Figure \ref{fig:AffineDquiver}.
The adjacency matrix can then be readily deduced from the figure.
\begin{equation}
	M^{\widehat{D}_r} = \left(\begin{array}{cccccccc}
		0 & 0 & 1 & 0 & \cdots & 0 & 0 & 0\\
		0 & 0 & 1 & 0 & \cdots & 0 & 0 & 0\\
		0 & 0 & 0 & 1 & 0 & \cdots & 0 & 0 \\
		\vdots & ~ & ~ & ~ & \ddots & ~ & ~ & ~\\
		0 & \cdots & 0 & 0 & 0 & 1 & 0 & 0 \\
		0 & \cdots & 0 & 0 & 0 & 0 & 1 & 1 \\
		0 & \cdots & 0 & 0 & 0 & 0 & 0 & 0 \\
		0 & \cdots & 0 & 0 & 0 & 0 & 0 & 0
	\end{array}\right)
\end{equation}
The labels $i,j$ run from $0$ (corresponding to the affine node) to $r$. The oriented $\widehat{E}_6$ quiver is shown in Figure \ref{fig:AffineE6quiver}.
The adjacency matrix is readily obtained to be
\begin{equation}
	M^{\widehat{E}_6} = \left(\begin{array}{ccccccc}
		0 & 0 & 1 & 0 & 0 & 0 & 0\\
		0 & 0 & 0 & 1 & 0 & 0 & 0\\
		0 & 0 & 0 & 0 & 1 & 0 & 0\\
		0 & 0 & 0 & 0 & 1 & 0 & 0\\
		0 & 0 & 0 & 0 & 0 & 1 & 0\\
		0 & 0 & 0 & 0 & 0 & 0 & 1\\
		0 & 0 & 0 & 0 & 0 & 0 & 0
	\end{array}\right).
\end{equation}
The labels $i,j$ run from $0$ (affine node) to $6$. In the $\widehat{E}_7$ case the quiver diagram with the corresponding labellings is shown in figure \ref{fig:AffineE7quiver}.
The adjacency matrix in this case is given by
\begin{equation}
	M^{\widehat{E}_7} = \left(\begin{array}{cccccccc}
		0 & 1 & 0 & 0 & 0 & 0 & 0 & 0\\
		0 & 0 & 0 & 1 & 0 & 0 & 0 & 0\\
		0 & 0 & 0 & 0 & 1 & 0 & 0 & 0\\
		0 & 0 & 0 & 0 & 1 & 0 & 0 & 0\\
		0 & 0 & 0 & 0 & 0 & 1 & 0 & 0\\
		0 & 0 & 0 & 0 & 0 & 0 & 1 & 0\\
		0 & 0 & 0 & 0 & 0 & 0 & 0 & 1\\
		0 & 0 & 0 & 0 & 0 & 0 & 0 & 0
	\end{array}\right).
\end{equation}
The labels $i,j$ run from $0$ (affine node) to $7$. The quiver for the $\widehat{E}_8$ case is shown in figure \ref{fig:AffineE8quiver}.
The adjacency matrix is given by
\begin{equation}
	M^{\widehat{E}_8} = \left(\begin{array}{ccccccccc}
		0 & 0 & 0 & 0 & 0 & 0 & 0 & 0 & 1\\
		0 & 0 & 0 & 1 & 0 & 0 & 0 & 0 & 0\\
		0 & 0 & 0 & 0 & 1 & 0 & 0 & 0 & 0\\
		0 & 0 & 0 & 0 & 1 & 0 & 0 & 0 & 0\\
		0 & 0 & 0 & 0 & 0 & 1 & 0 & 0 & 0\\
		0 & 0 & 0 & 0 & 0 & 0 & 1 & 0 & 0\\
		0 & 0 & 0 & 0 & 0 & 0 & 0 & 1 & 0\\
		0 & 0 & 0 & 0 & 0 & 0 & 0 & 0 & 1\\
		0 & 0 & 0 & 0 & 0 & 0 & 0 & 0 & 0	
	\end{array}\right).
\end{equation}
$i$ and $j$ run from $0$ (affine node) to $8$. In all the quivers, $d_i$'s are fixed by the anomaly condition (\ref{eq:anoCond}) which can be rewritten in terms of $M$,
\begin{equation}
2d_i = \sum_{j} (M_{ij} + M_{ji}) d_j.
\end{equation}

\subsection{Elliptic Genus}

Let us now come to the computation of the elliptic genus. In order to proceed we fix our notation for the theta-function on which the elliptic genus depends
\begin{equation} \label{eq:originaltheta}
	\theta(x;q_{\tau}) = i q_{\tau}^{1/8} x^{1/2} \prod_{k=1}^{\infty} (1-q_{\tau}^k)(1-q_{\tau}^k x)(1-q_{\tau}^{k-1} x^{-1}),
\end{equation}
where we have defined
\begin{equation}
	q_{\tau} := e^{2\pi i \tau}, \quad x := e^{2\pi i z}.
\end{equation}
In the following we will neglect the explicit dependence on $q_{\tau}$ in the theta-function and just write $\theta(x)$. Building on earlier work \cite{Haghighat:2013gba,Haghighat:2013tka}, the elliptic genera of strings of ADE quiver theories have been computed in \cite{Gadde:2015tra}. The elliptic genera of a collection of strings with gauge charges $k_i$, with $i$ corresponding to an inner node of the quiver, can be characterised in terms of Young diagrams $Y_l^{(i)}, l = 1, \ldots, N d_i$. As $i=0$ corresponds to an affine node not present in the inner quiver it is assumed that $k_0 = 0$ and hence $Y_l^{(0)} = \emptyset$. In the $A_r$ case we also have $Y_l^{(r+1)} = \emptyset$. The number of boxes $n^{(i)}_l$ in $Y_l^{(i)}$ obey $\sum_{l=1}^{N d_i} n_l^{(i)}  = k_i$. For ease of notation we will set $N=1$ in the following. With these conventions and the definition of adjacency matrices above, the elliptic genus of the general $ADE$ case as a function of the chemical potentials $a_l^{(i)}$ , $q=e^{\epsilon_1}$ and $t=e^{-\epsilon_2}$ can be expressed as\footnote{For $A$ type quiver, it is understood that the rank $r$ should be replaced by $r+1$.}
\begin{eqnarray}
	Z_{\vec{k}} & = & \sum_{\{Y_l^{(i)}\}} \prod_{i=0}^r \prod_{l,m=1}^{d_i} \prod_{\stackrel{(x_1,y_1) \in Y_l^{(i)}}{(x_2,y_2) \in Y_m^{(i)}}} \left(\frac{\theta(\frac{a_l^{(i)}}{a_m^{(i)}}q^{x_2^{(i,m)}-x_1^{(i,l)}}t^{y_1^{(i,l)}-y_2^{(i,m)}})\theta(\ada{i}{l}{i}{m} q^{\xb{2}{i}{m}-\xb{1}{i}{l}-1}t^{\yb{1}{i}{l}-\yb{2}{i}{m}+1})}{\theta(\frac{a_l^{(i)}}{a_m^{(i)}}q^{x_2^{(i,m)}-x_1^{(i,l)}-1}t^{y_1^{(i,l)}-y_2^{(i,m)}})\theta(\ada{i}{n}{i}{m} q^{\xb{2}{i}{m}-\xb{1}{i}{l}}t^{\yb{1}{i}{l}-\yb{2}{i}{m}+1})}\right) \nonumber \\
	~ & \times &  \prod_{i,j=0}^r \prod_{l=1}^{d_i} \prod_{m=1}^{d_j} \prod_{\stackrel{(x_1,y_1)\in Y^{(i)}_l}{(x_2,y_2) \in Y^{(j)}_m}} \left(\frac{\theta(\ada{i}{l}{j}{m}q^{\xb{2}{j}{m}-\xb{1}{i}{l}+\half}t^{\yb{1}{i}{l}-\yb{2}{j}{m}+\half})\theta(\ada{i}{l}{j}{m}q^{\xb{2}{j}{m}-\xb{1}{i}{l}-\half}t^{\yb{1}{i}{l}-\yb{2}{j}{m}-\half})}{\theta(\ada{i}{l}{j}{m}q^{\xb{2}{j}{m}-\xb{1}{i}{l}+\half}t^{\yb{1}{i}{l}-\yb{2}{j}{k}-\half})\theta(\ada{i}{l}{j}{m}q^{\xb{2}{j}{m}-\xb{1}{i}{l}-\half}t^{\yb{1}{i}{l}-\yb{2}{j}{m}+\half})}\right)^{M_{ij}} \nonumber \\
~ & \times &  \prod_{i=0}^r \prod_{l=1}^{d_i} \prod_{(x,y)\in Y^{(i)}_l}\left( \frac{\prod_{j=0}^r\prod_{m=1}^{d_j}\theta(\ada{i}{l}{j}{m}q^{-x-\half}t^{y+\half})^{M_{ij}}\theta(\ada{j}{m}{i}{l}q^{x+\half}t^{-y-\half})^{M_{ji}}}{\prod_{m=1}^{d_i}\theta(\ada{i}{l}{i}{m}q^{-x-1}t^{y+1})\theta(\ada{i}{m}{i}{l}q^x t^{-y})}\right)  \label{eq:Z}
\end{eqnarray}
Some comments are at order here. An important difference between the $A$ case as compared to the $D$ and $E$ cases is that in the former case one can refine the index with respect to an extra $U(1)$ symmetry which we shall denote by $U(1)_{\sigma}$. Such a refinement is not possible for the $D$ and $E$ cases as $U(1)_{\sigma}$ becomes anomalous. This is due to the fact that $A$ type ALE spaces have an extra $U(1)$ isometry as compared to the $D$ and $E$ cases. For reasons which will become clear soon, we shall call such a refinement \textit{mass-deformation}. Strictly speaking, equation (\ref{eq:Z}) is only valid in the mass-less $\sigma=0$ limit. Therefore, in the $A$ case, one has to do the following substitution
\begin{equation}
	\ada{i}{l}{j}{m} \longrightarrow \ada{i}{l}{j}{m} e^{2\pi i \sigma}, \quad \textrm{whenever} \quad i \neq j.
\end{equation}
Moreover, in this paper we will mainly work in the unrefined limit $\epsilon_1 = -\epsilon_2 = \hbar$ or in other words $q = t$.

Let us now proceed to simplify equation (\ref{eq:Z}). Using the identity
\begin{eqnarray}
	~ & ~ & \prod_{\stackrel{(x_1,y_2) \in \nu}{(x_2,y_2)\in \mu}} \frac{\theta(Qq^{y_1-y_2}t^{x_2-x_1+1})\theta(Q q^{y_1-y_2-1}t^{x_2-x_1})}{\theta(Q q^{y_1-y_2}t^{x_2 - x_1})\theta(Q q^{y_1-y_2-1}t^{x_2-x_1+1})} \nonumber \\
	~ & = & \left(\prod_{(x,y)\in \nu}\frac{\theta(Q q^{\nu_i-j}t^{\mu^t_j-i+1})}{\theta(Q q^{\nu_i - j }t^{-i+1})}\right) \left(\prod_{(x,y)\in \mu}\frac{\theta(Q q^{-\mu_i+j-1}t^{-\nu^t_j+i})}{\theta(Q q^{-\mu_i+j-1}t^i)}\right),\nonumber
\end{eqnarray}
and its unrefined version
\begin{eqnarray}
	~ & ~ & \prod_{\stackrel{(x_1,y_2) \in \nu}{(x_2,y_2)\in \mu}} \frac{\theta(Qq^{y_1-y_2+x_2-x_1+1})\theta(Q q^{y_1-y_2-1+x_2-x_1})}{\theta(Q q^{y_1-y_2+x_2 - x_1})\theta(Q q^{y_1-y_2+x_2-x_1})} \nonumber \\
	~ & = & \left(\prod_{(x,y)\in \nu}\frac{\theta(Q q^{\nu_i-j+\mu^t_j-i+1})}{\theta(Q q^{\nu_i - j-i+1})}\right) \left(\prod_{(x,y)\in \mu}\frac{\theta(Q q^{-\mu_i+j-\nu^t_j+i-1})}{\theta(Q q^{-\mu_i+j-1+i})}\right),\nonumber
\end{eqnarray}	
we can rewrite the first two lines of (\ref{eq:Z}) as follows. The first line becomes after passing to the unrefined limit $t=q$:
\begin{eqnarray}
	~ & ~ & \left.\prod_{\stackrel{(x_1,y_1) \in Y_l^{(i)}}{(x_2,y_2) \in Y_m^{(i)}}} \left(\frac{\theta(\frac{a_l^{(i)}}{a_m^{(i)}}q^{x_2^{(i,m)}-x_1^{(i,l)}}t^{y_1^{(i,l)}-y_2^{(i,m)}})\theta(\ada{i}{l}{i}{m} q^{\xb{2}{i}{m}-\xb{1}{i}{l}-1}t^{\yb{1}{i}{l}-\yb{2}{i}{m}+1})}{\theta(\frac{a_l^{(i)}}{a_m^{(i)}}q^{x_2^{(i,m)}-x_1^{(i,l)}-1}t^{y_1^{(i,l)}-y_2^{(i,m)}})\theta(\ada{i}{n}{i}{m} q^{\xb{2}{i}{m}-\xb{1}{i}{l}}t^{\yb{1}{i}{l}-\yb{2}{i}{m}+1})}\right)\right|_{t=q} \nonumber \\
	~ & = & \left(\prod_{(x,y) \in Y^{(i)}_m} \frac{\theta(\ada{i}{l}{i}{m} q^{x-y})}{\theta(\ada{i}{l}{i}{m} q^{\nu^{(i,m)}_y-x+\nu^{(i,l),t}_x-y+1})}\right)\left(\prod_{(x,y)\in Y^{(i)}_l} \frac{\theta(\ada{i}{l}{i}{m} q^{y-x})}{\theta(\ada{i}{l}{i}{m}q^{-\nu^{(i,l)}_y+x-1 - \nu^{(i,m)}_x+y})}\right), \nonumber \\ \label{eq:Z1}
\end{eqnarray}
and the second line becomes
\begin{eqnarray}
	~ & ~ & \left.\prod_{\stackrel{(x_1,y_1)\in Y^{(i)}_l}{(x_2,y_2) \in Y^{(j)}_m}} \left(\frac{\theta(\ada{i}{l}{j}{m}q^{\xb{2}{j}{m}-\xb{1}{i}{l}+\half}t^{\yb{1}{i}{l}-\yb{2}{j}{m}+\half})\theta(\ada{i}{l}{j}{m}q^{\xb{2}{j}{m}-\xb{1}{i}{l}-\half}t^{\yb{1}{i}{l}-\yb{2}{j}{m}-\half})}{\theta(\ada{i}{l}{j}{m}q^{\xb{2}{j}{m}-\xb{1}{i}{l}+\half}t^{\yb{1}{i}{l}-\yb{2}{j}{k}-\half})\theta(\ada{i}{l}{j}{m}q^{\xb{2}{j}{m}-\xb{1}{i}{l}-\half}t^{\yb{1}{i}{l}-\yb{2}{j}{m}+\half})}\right)^{M_{ij}}\right|_{t=q} \nonumber \\
	~ & = & \left(\prod_{(x,y)\in Y^{(j)}_m} \frac{\theta(\ada{i}{l}{j}{m}q^{\nu^{(j,m)}_y - x + \nu^{(i,l),t}_x-y+1})}{\theta(\ada{i}{l}{j}{m} q^{x-y})}\right)^{M_{ij}}\left(\prod_{(x,y) \in Y^{(i)}_l} \frac{\theta(\ada{i}{l}{j}{m} q^{-\nu^{(i,l)}_y + x - 1 - \nu^{(j,m),t}_x + y})}{\theta(\ada{i}{l}{j}{m} q^{y-x})}\right)^{M_{ij}}. \label{eq:Z2}
\end{eqnarray}
Furthermore, taking the unrefined limit of the third line of equation (\ref{eq:Z}) we obtain
\begin{eqnarray}
	~ & ~ & \left.\prod_{(x,y)\in Y^{(i)}_l}\left( \frac{\prod_{j=0}^r\prod_{m=1}^{d_j}\theta(\ada{i}{l}{j}{m}q^{-x-\half}t^{y+\half})^{M_{ij}}\theta(\ada{j}{m}{i}{l}q^{x+\half}t^{-y-\half})^{M_{ji}}}{\prod_{m=1}^{d_i}\theta(\ada{i}{l}{i}{m}q^{-x-1}t^{y+1})\theta(\ada{i}{m}{i}{l}q^x t^{-y})}\right)\right|_{t=q} \nonumber \\
	~ & = & \prod_{(x,y)\in Y^{(i)}_l}\left( \frac{\prod_{j=0}^r\prod_{m=1}^{d_j}\theta(\ada{i}{l}{j}{m}q^{y-x})^{M_{ij}}\theta(\ada{j}{m}{i}{l}q^{x-y})^{M_{ji}}}{\prod_{m=1}^{d_i}\theta(\ada{i}{l}{i}{m}q^{y-x})\theta(\ada{i}{m}{i}{l}q^{x-y})}\right) \label{eq:Z3}
\end{eqnarray}
Combining all three equations (\ref{eq:Z1}), (\ref{eq:Z2}) and (\ref{eq:Z3}) we see that there are many cancellations and the final form of $Z$ in the unrefined limit becomes
\begin{eqnarray}
	Z_{\vec{k}} & = & \sum_{\{Y_l^{(i)}\}} \prod_{i=0}^r \prod_{l,m=1}^{d_i} \left(\prod_{(x,y) \in Y^{(i)}_m} \frac{1}{\theta(\ada{i}{l}{i}{m} q^{\nu^{(i,m)}_y-x+\nu^{(i,l),t}_x-y+1})}\right)\left(\prod_{(x,y)\in Y^{(i)}_l} \frac{1}{\theta(\ada{i}{l}{i}{m}q^{-\nu^{(i,l)}_y+x-1 - \nu^{(i,m)}_x+y})}\right) \nonumber \\
	~ &  \times & \prod_{i,j=0}^r \prod_{l=1}^{d_i} \prod_{m=1}^{d_j}  \left(\prod_{(x,y)\in Y^{(j)}_m} \theta(\ada{i}{l}{j}{m}q^{\nu^{(j,m)}_y - x + \nu^{(i,l),t}_x-y+1})^{M_{ij}}\right)\left(\prod_{(x,y) \in Y^{(i)}_l} \theta(\ada{i}{l}{j}{m} q^{-\nu^{(i,l)}_y + x - 1 - \nu^{(j,m),t}_x + y})^{M_{ij}}\right). \nonumber \\
 \label{eq:Zfinal}
\end{eqnarray}
One can now check that the modular anomaly of (\ref{eq:Zfinal}) matches the one given in (\ref{eq:anomaly}). To see this, one has to use the following identity between partition sums
\begin{equation}
	\sum_{k,l=1}^N \sum_{(x,y) \in Y_k} (h_{k,l}(x,y)^2 - (y-x)^2) = |Y|^2,
\end{equation}
where we are using the definitions
\begin{equation}
	h_{k,l}(x,y) = \nu_y^{(k)} - x + \nu_x^{(l),t} - y + 1, \quad |Y| = \sum_k |Y_k|.
\end{equation}

\subsection{Thermodynamic limit}
Next, we want to rewrite this result in terms of partition densities. In order to do this, we need a combinatoric Young tableaux identity which we state here without proof\footnote{$\sigma$ can be replaced by an arbitrary function here.}
\begin{equation} \label{eq:identity1}
	\prod_{x,y=1}^{\infty} \frac{\sigma(\nu_x - \mu_y + y - x)}{\sigma(y-x)} = \left(\prod_{(x,y) \in \nu}\frac{1}{\sigma(\nu_x - y + \mu^t_y - x +1)}\right)\left(\prod_{(x,y)\in \mu}\frac{1}{\sigma(-\mu_x + y + x - \nu^t_y - 1)}\right).
\end{equation}
Furthermore, we shall need the difference equation satisfied by the multi-gamma function $\gamma(z;\hbar)$
\begin{equation} \label{eq:identity2}
	2 \gamma(z;\hbar)  - \gamma(z + \hbar,\hbar) - \gamma(z-\hbar;\hbar)  = \ln \theta(z),
\end{equation}
where we refere to the appendix for further details on the functions $\gamma(z;\hbar)$. Now we can define partition function densities for each node $i=0, \cdots , r$ by setting
\begin{eqnarray}
	\varrho_i(z) & = & \sum_l 2 \delta(z - a^{(i)}_l) + 2 \sum_{x=1}^{\infty}\delta(z - \mu^{(i)}_l + \hbar (1 + \nu^{(i,l)}_x - x)) \nonumber \\
	~ & ~ & - \delta(z - \mu^{(i)}_l + \hbar(\nu^{(i,l)}_x - x)) - \delta(z - \mu^{(i)}_l + \hbar(1-x)) + \delta(z - \mu^{(i)}_l - \hbar x). \label{eq:partitiondensity}
\end{eqnarray}
We note here that the support of each density function is given by a collection of intervals $I^{(i)}_l$, $l=1, \ldots, d_i$ and we have
\begin{equation}
	\mu^{(i)}_l = \int_{I^{(i)}_l} z \varrho_i(z) dz, \quad \int_{I^{(i)}_l} \varrho_i(z) dz = 2,
\end{equation}
while for the affine node we impose
\begin{equation}
	\varrho_0(z) = \sum_l 2 \delta(z - \mu^{(0)}_l).
\end{equation}
In the $A$ type case we have to add the condition
\begin{equation}
	\varrho_{r+1}(z) = \sum_l 2 \delta(z-\mu^{(r+1)}_l).
\end{equation}
Furthermore, the string charges are determined by the identity
\begin{equation}
	k_i = -\frac{1}{2\hbar^2} \sum_{l=1}^{d_i} {\mu^{(i)}_l}^2 + \frac{1}{4 \hbar^2} \int dz z^2 \varrho_i(z).
\end{equation}
Now, given an affine quiver $\widehat{\mathrm{q}}$, we are in the position to use identities (\ref{eq:identity1}) and (\ref{eq:identity2}) to rewrite $Z_{\vec{k}}$ given in (\ref{eq:Zfinal}) as follows:
\begin{eqnarray}
	Z_{\vec{k}} & = & \exp\left( - \frac{1}{4} \int_{\mathbb{R}^2} dz' dz'' \sum_{i \in \textrm{Vert}_{\widehat{\mathrm{q}}}} \varrho_i(z') \varrho_i(z'') \gamma(z' - z'';\hbar)\right. \nonumber \\
	~ & ~ & + \left. \frac{1}{4} \int_{\mathbb{R}^2} dz' dz'' \sum_{e \in \textrm{Edge}_{\widehat{\mathrm{q}}}} \varrho_{t(e)}(z') \varrho_{s(e)}(z'') \gamma(z' - z''+\sigma;\hbar)\right). \label{eq:Zfinal2}
\end{eqnarray}
In the above we have used notation from reference \cite{Nekrasov:2012xe}. Also, comparing to \cite{Nekrasov:2012xe}, one sees that the parameter $\sigma$ corresponds to the mass of bi-fudamental hypermultiplets connecting adjacent nodes, hence its interpretation as \textit{mass-deformation}. In the following we shall set $\sigma = 0$ in order to keep a uniform notation for the $A$, $D$ and $E$ quivers. Then in the thermodynamic limit the full partition function is approximated by
\begin{equation}
	Z_{T^2 \times \mathbb{R}^4} \equiv \int \prod_i \mathcal{D} \varrho_i \prod_{i,l} \mu^{D (i)}_l \exp\left[\frac{1}{2\hbar^2}\mathcal{F}_0 + \mathcal{O}(\hbar)\right],
\end{equation}
with $\mathcal{F}_0$ given by
\begin{eqnarray}
	\mathcal{F}_0 & = & \sum_{i \in \textrm{Vert}_{\mathrm{q}}}2\pi i \rho_i  \int_{\mathbb{R}} dz \varrho_i(z) \frac{z^2}{2} - \frac{1}{2} \int_{\mathbb{R}^2} dz' dz'' \sum_{i \in \textrm{Vert}_{\widehat{\mathrm{q}}}} \varrho_i(z') \varrho_i(z'') \gamma_0(z' - z'') \nonumber \\
	~ & ~ & + \frac{1}{2} \int_{\mathbb{R}^2} dz' dz'' \sum_{e \in \textrm{Edge}_{\widehat{\mathrm{q}}}} \varrho_{t(e)}(z') \varrho_{s(e)}(z'') \gamma_0(z' - z'') + \sum_{i,l} \mu^{D (i)}_l \left(\mu^{(i)}_l - \int_{I_{i,l}} z \varrho_i(z) dz \right). \nonumber \\ \label{eq:affineF0}
\end{eqnarray}
In order to derive the above result, we have used the following expansion of the elliptic multiple gamma function
\begin{equation}
	\gamma(z;\hbar)=\sum_{g=0}^{\infty}\hbar^{2g-2}\gamma_g(z), \quad \gamma_0(z) = \log \theta(z).
\end{equation}
This result can be re-expressed in terms of the ordinary quiver and the contributions coming from the affine/boundary nodes
\begin{eqnarray}
	\mathcal{F}_0 & = & \sum_{i \in \textrm{Vert}_{\mathrm{q}}}2\pi i \rho_i  \int_{\mathbb{R}} dz \varrho_i(z) \frac{z^2}{2} - \frac{1}{2} \int_{\mathbb{R}^2} dz' dz'' \sum_{i \in \textrm{Vert}_{\mathrm{q}}} \varrho_i(z') \varrho_i(z'') \gamma_0(z' - z'') \nonumber \\
	~ & ~ & + \frac{1}{2} \int_{\mathbb{R}^2} dz' dz'' \sum_{e \in \textrm{Edge}_{\mathrm{q}}} \varrho_{t(e)}(z') \varrho_{s(e)}(z'') \gamma_0(z' - z'') + \sum_{i,l} \mu^{D (i)}_l \left(\mu^{(i)}_l - \int_{I_{i,l}} z \varrho_i(z) dz \right)\nonumber \\
	~ & ~ & +\frac{1}{2}\sum_{i \in \textrm{b}_{\mathrm{q}}} \int_{\mathbb{R}} dz \varrho_i(z) \sum_l  \gamma_0(z - m^{(i)}_l), \nonumber \\ \label{eq:F0}
\end{eqnarray}
where terms independent of $\varrho_i$ have been omitted. We have introduced the notation $\textrm{b}_{\mathrm{q}}$ for the \textit{boundary}-nodes of the quiver $\mathrm{q}$. For the $A_r$ quiver the set $\textrm{b}_{\mathrm{q}}$ includes the nodes $i=1$ and $i=r$, whereas for the $D$ and $E$ type quivers this set contains the node adjacent to the affine node of the corresponding quiver. Furthermore, we have introduced $m_l^{(1)} = \mu_l^{(0)}$ and $m_l^{(r)} = \mu_l^{(r+1)}$ in the case of the $A_r$-quiver, whereas in the case of the $D$ and $E$ quivers $m_l^{(i)}=\mu_l^{(0)}$ ($i$ being the single element in $\textrm{b}_{\mathrm{q}}$). The result (\ref{eq:F0}) is similar to the expression for the prepotential presented in section 5.2 of \cite{Nekrasov:2012xe}. It is in fact the elliptic version of the quoted expression which gives the result for purely four-dimensional quiver gauge theories where the volume of the $T^2$ in the compactification from six to four dimensions is taken to zero. Therefore, our result generalizes the work of \cite{Nekrasov:2012xe} to the setting with non-trivial size for $T^2$ where we focus on their type I case in this section. From this correspondence one can also see that in our setup we are getting fundamental matter for each edge connecting a boundary node to an affine node with masses given by $m_l^{(i)}$.

\subsection{Seiberg-Witten geometry}
Variation with respect to $\varrho_i(z)$ with $z \in I_{i,l}$ then leads to the following saddle point equation
\begin{eqnarray}
 z a^{D (i)}_l & = &	- \int_{\mathbb{R}} dz' \sum_{i \in \textrm{Vert}_{\mathrm{q}}} \varrho_i(z') \gamma_0(z-z') + \frac{1}{2} \int_{\mathbb{R}} dz' \sum_j \varrho_j(z') (M + M^T)_{ij}\gamma_0(z-z') \nonumber \\
 ~ & ~ & + \frac{z^2}{2} 2\pi \rho_i + \frac{1}{2} \sum_{k \in b_{\mathrm{q}}} \sum_{l'=1}^N \gamma_0(z-m_{l'}^{(k)}) \delta_{ik}.
\end{eqnarray}
We can rewrite the second derivative of the above equation with respect to $z$ as a non-linear polynomial difference equation by exponentiation:
\begin{equation} \label{eq:iWeyl}
	y_i^+(z) y^-_i(z) = \mathcal{P}_i \prod_j y_j(z)^{(M+M^T)_{ij}}~, \quad \mathcal{P}_i =e^{2\pi i \rho_i} \prod_{k \in \textrm{b}_{\mathrm{q}}} \prod_{l'=1}^N \theta(z- m_{l'}^{(k)})^{\delta_{ik}}~,
\end{equation}
where we define
\begin{equation}
	y_i(z) = \exp \frac{1}{2} \int_{\mathbb{R}} dz' \varrho_i(z') \log \theta_1(z-z'),
\end{equation}
for $z \in I_{i,l}$, $l=1, \ldots, N d_i$, and the following notation is implied
\begin{equation}
y^{\pm}_i(z) = y_i(z \pm i 0), \quad z \in I_{i,l}.
\end{equation}
Equation (\ref{eq:iWeyl}) can be interpreted as a Weyl transformation upon crossing the cuts $I_i$ \cite{Nekrasov:2012xe}. In order to describe the Seiberg-Witten curve we thus have to find the Weyl-invariant combinations of the $y_i(z)$ functions. Such a construction is known as the spectral curve construction and has been described in \cite{Nekrasov:2012xe}. We will follow that reference in the following exposition where we shall be brief. Let $\mathbf{G} = \mathbf{G}_{\mathrm{q}}$ be the simple complex Lie group corresponding to non-affine ADE quivers. Suppose $\lambda$ is a dominant weight, i.e. $\lambda(\alpha_i^{\vee}) \geq 0$ for all $i \in \textrm{Vert}_{\mathrm{q}}$. Let $R_{\lambda}$ be the irreducible highest weight module of $\mathbf{G}_{\mathrm{q}}$ with highest weight $\lambda$, with corresponding homomorphism given by $\pi_{\lambda} : \mathbf{G}_{\mathrm{q}} \rightarrow \textrm{End}(R_{\lambda})$. Then, defining the torus
\begin{equation}
	\mathbb{T}_{\langle z \rangle} \equiv \mathbb{C}_{\langle z \rangle}/\{\mathbb{Z} + \tau \mathbb{Z}\},
\end{equation}
the spectral curve in $\mathbb{T}_{\langle z \rangle} \times \mathbb{C}_{\langle t \rangle}$ is
\begin{equation}
	\textrm{det}_{R_{\lambda}} \left(1 - t^{-1} \zeta(z)^{-1} \pi_{\lambda}(g(x))\right) = 0, \quad g(x) \in G_{\mathrm{q}}, \quad x = e^{2\pi i z},
\end{equation}
where $\zeta(z)$ is a normalization constant which we shall not specify further here. To finish the present discussion, we also need to specify the Seiberg-Witten differential. The above curve comes with a canonical differential, which is the restriction of the following differential form on $\mathbb{T}_{\langle z \rangle} \times \mathbb{C}_{\langle t \rangle}$:
\begin{equation}
	\lambda = x \frac{dt}{t}.
\end{equation}

We shall now describe the $A_r$ case in some more detail. We will focus on the representation $R_{\lambda_1}$ where $\lambda_1$ is the first fundamental weight. The corresponding group element $g_{\lambda_1}(z)$ is the diagonal matrix
\begin{equation}
	g_{\lambda_1}(z) = \textrm{diag}(t_1(z), \ldots, t_{r+1}(z))
\end{equation}
with
\begin{eqnarray}
	~ & ~ & t_1(z) =\zeta(z) y_1(z), \quad t_{r+1}(z) = \zeta(z) \mathcal{P}^{[r]} y_r(z)^{-1}, \nonumber \\
	~ & ~ & t_i(z) = \zeta(z) \mathcal{P}^{[i-1]}(z) y_i(z) y_{i-1}(z)^{-1}, \quad i=2,\ldots,r,
\end{eqnarray}
where we are using the notation
\begin{equation}
	\mathcal{P}^{[i]} = \prod_{k=1}^{i} \mathcal{P}_k.
\end{equation}
Next, we shall need the fundamental characters $\chi_i$ which are the characters of the representations $\Lambda^i \mathbb{C}^{r+1}$. The $\chi_i$ are invariants under Weyl transformations of the $y_i$ and are given by
\begin{equation}
	\chi_i(y(z)) = \prod_{j=1}^{i-1} \mathcal{P}_j^{j-i} e_i(y_1,y_2 y_1^{-1} \mathcal{P}^{[1]}, \ldots, y_i y_{i-1}^{-1} \mathcal{P}^{[i-1]}, \ldots, y_r^{-1} \mathcal{P}^{[r]}),
\end{equation}
where the $e_i$ are elementary symmetric polynomials in $r+1$ variables. Using these definitions, it can be shown that the spectral curve equation becomes
\begin{eqnarray}
	\textrm{det}(t \cdot 1_{r+1} - g_{\lambda_1}(z)) & = & t^{r+1} + \sum_{i=1}^r (-1)^i t^{r+1-i} \zeta(z)^i \prod_{j=1}^{i-1} \mathcal{P}_j^{i-j}(z) \chi_i(y(z)) \nonumber \\
	~ & ~ & + (-\zeta(z))^{r+1} \prod_{j=1}^r \mathcal{P}_j^{r+1-j}(z). \label{eq:Arspc}
\end{eqnarray}
Up to this point we have been following the exposition of \cite{Nekrasov:2012xe} except for the fact that the spectral curve is now an $r+1$-fold cover of $\mathbb{T}_{\langle z \rangle}$ instead of $\mathbb{C}_{\langle z \rangle}$. This has a significant impact on the master equations of \cite{Nekrasov:2012xe}. In order to evaluate (\ref{eq:Arspc}), we have to evaluate the functions $\chi_i(y(z))$ on the torus $\mathbb{T}_{\langle z \rangle}$. In our case these functions are living in the determinant bundle of a flat $SL(N d_i)_{\mathbb{C}}$ bundle on $\mathbb{T}_{\langle z \rangle}$. As such they are sections of line bundles $L_i$ on $\mathbb{T}_{\langle z \rangle}$. These line bundles are specified in terms of divisors on $\mathbb{T}_{\langle z \rangle}$
\begin{equation}
	D_L =  z_1 + \cdots + z_{N d_i},
\end{equation}
where $\sum_{l} z_l = z_0$ and $z_0$ is the point corresponding to the identity of the abelian group law on the elliptic curve $\mathbb{T}_{\langle z \rangle}$. In the $A_r$ case all $d_i = 1$ and so the corresponding sections are of degree $N$ and are given by
\begin{equation} \label{eq:si}
	s^f_i(z;\mathbf{z}) = T_{i,0} \prod_{l=1}^N \theta(x/x_{i,l};q_{\tau}), \quad x=e^{2\pi i z}, ~x_{i,l} = e^{2\pi z_{i,l}},
\end{equation}
where the $z_{i,l}$ are functions of the gauge fugacities, masses and gauge couplings $\rho_i$, and $T_{i,0}$ is a normalization constant. Furthermore, in (\ref{eq:si}) we are using a slightly modified version of the theta-function given by
\begin{equation} \label{eq:theta}
	\theta(x;q) = \prod_{n\geq 0} (1-x q^n)(1-q^{n+1})(1-x^{-1}q^{n+1}),
\end{equation}
in order to obtain the right limiting behavior in the $q_{\tau} \rightarrow 0$ limit:
\begin{equation}
	\lim_{q_{\tau}\rightarrow 0} s^f_i = T_{i,0} \prod_{l=1}^N x_l^{-1}(x_l-x) = \frac{T_{i,0}}{x_1 \cdots x_N} (-x)^N + \cdots = T_{i,0} (-x)^N + \mathcal{O}(x^{N-1}).
\end{equation}
This is the correct leading behavior for the functions $\chi_i(y(z))$ in the purely four-dimensional case of the $A$-type quiver discussed in \cite{Nekrasov:2012xe}, where it is understood that the constants $T_{i,0}$ are given by
\begin{equation}
	T_{i,0} = \left(\prod_{j=1}^{i-1} q_{\rho_j}^{j-1}\right) e_i(1,q_{\rho_1},q_{\rho_1}q_{\rho_2},\ldots,q_{\rho_1} \cdots q_{\rho_r}), \quad q_{\rho_i} = e^{2\pi i \rho_i},
\end{equation}
and $e_i$ are the elementary symmetric polynomials in $r+1$ variables. Altogether, we thus arrive at the master equations
\begin{equation} \label{eq:fibersection}
	\chi_i(y(z)) = s^f_i(z;\mathbf{z}), \quad z_i = z_i(\boldsymbol{\mu},\mathbf{m},\boldsymbol{\rho}).
\end{equation}
The same master equations also hold for the $D-$ and $E-$cases with the only difference being that the degrees of the sections $s_i$ will differ since the Dynkin labels $d_i$ are not all equal. Returning to the $A_r$ case, note that $\mathcal{P}_1$ and $\mathcal{P}_r$ are sections of degree $N$ line bundles whereas $\mathcal{P}_i$ for $1 < i < r$ are of degree $0$. Taking the variable $t$ to be a section of degree $N$ we thus see that equation (\ref{eq:Arspc}) is of degree $(r+1)N$. 

\section{Mirror Symmetry}
\label{sec:mirrorsymmetry}

In this section we want to connect our results to the work of \cite{Kanazawa:2016tnt} on SYZ mirror symmetry and give a physics interpretation for their geometric setup. The fundamental building block in the constructions of \cite{Kanazawa:2016tnt} is a local Calabi-Yau surface of type $\widehat{A}_{d-1}$ for $d \geq 1$. This space is the total space of the elliptic fibration over the unit disc $\mathbb{D} = \{|z| < 1\} \subset \mathbb{C}$, where all fibers are smooth except for the central fiber, which is a nodal union of $d$ rational curves forming a cycle. This surface geometry has a natural extension to higher dimensions as follows. For $(d_1,\ldots,d_{n-1}) \in \mathbb{Z}^{n-1}_{\geq 1}$, one can consider the multiple fiber product $\widehat{A}_{d_1-1} \times_{\mathbb{D}} \ldots \times_{\mathbb{D}} \widehat{A}_{d_{n-1}-1}$, giving rise to a local Calabi-Yau $n$-fold of type $\widehat{A}$. In this paper we will be interested in the case $n=3$ giving rise to a Calabi-Yau three-fold $X_{d_1,d_2}$ which admits two different elliptic fibrations:
\begin{equation}
	\pi_1 : X_{(d_1,d_2)} \rightarrow \widehat{A}_{d_1-1}, \quad \pi_2 : X_{(d_1,d_2)} \rightarrow \widehat{A}_{d_2-1}.
\end{equation}
The corresponding two different elliptic fibers can be seen as fiber and base of the local Calabi-Yau geometry. Since the choice of fiber and base is arbitrary, this gives rise to the so called \textit{fiber-base} duality as explored in \cite{Katz:1997eq}. In the following we will take the overall volume of the base elliptic cure to be $\rho$ and the one of the fiber to be $\tau$:
\begin{equation}
	\textrm{Vol}(T^2_{\textrm{base}}) = \rho, \quad \textrm{Vol}(T^2_{\textrm{fiber}}) = \tau,  \quad q_{\rho} = e^{2\pi i \rho}, \quad q_{\tau} = e^{2\pi \tau}.
\end{equation}
Let us now focus on the case relevant for our paper, namely $d_1 = r+1$ and $d_2 = N$. Then, as shown in \cite{Kanazawa:2016tnt}, the SYZ mirror of $X_{(r+1,N)}$ is
\begin{equation} \label{eq:CYeq}
	uv = F^{\textrm{open}},
\end{equation}
where $F^{\textrm{open}}$ is the open Gromov-Witten potential given by\footnote{Our variables $z_1$ and $z_2$ are shifted by overall constants $\frac{1}{p}\sum_{k=0}^{p-1} k \tau_({-1-k,0)}$ and $\frac{1}{q}\sum_{l=0}^{q-1}l \rho_{(0,-1-l)}$ as compared to Theorem 5.9 of \cite{Kanazawa:2016tnt}.}
\begin{equation} \label{eq:mirrorcurve}
	\sum_{i=0}^r \sum_{l=0}^{N-1} K_{i,l} \Delta_{i,l} \Theta_2\left[\begin{array}{c}(\frac{i}{r+1},\frac{l}{N})\\(\frac{-(r+1)\rho}{2},-\frac{N\tau}{2})\end{array}\right]\left((r+1)z_1,N z_2; \left[\begin{array}{cc}(r+1)\rho & \sigma \\ \sigma & N \tau\end{array}\right]\right).
\end{equation}
Let us explain the notation above. $\Delta_{i,l}$ are open Gromov-Witten generating functions and $K_{i,l}$ are given by\footnote{We are including the $\tau_{(-1-k,0)}$ and $\rho_{(0,-1-l)}$ dependent factors in the definition of $\Delta_{i,l}$ as compared to Theorem 5.9 of \cite{Kanazawa:2016tnt}.}
\begin{equation}
	K_{i,l} = q_{\rho}^{\frac{i}{2} - \frac{i^2}{2(r+1)}} q_{\tau}^{\frac{l}{2}-\frac{l^2}{2N}}.
\end{equation}
Last but not least, the theta function in (\ref{eq:mirrorcurve}) is the genus $2$ theta function. The genus $g$ theta function is defined as follows
\begin{equation}
	\Theta_g\left[\begin{array}{c}
	\vec{a}\\
	\vec{b}
	\end{array}\right](\vec{z};\Omega) = \sum_{\vec{n} \in \mathbb{Z}^g} \exp\left(\frac{1}{2}(\vec{n} + \vec{a})^t \Omega (\vec{n}+\vec{a}) + (\vec{n} + \vec{a})\cdot (\vec{z} + \vec{b})\right),
\end{equation}
where $\Omega$ is an element of the Siegel upper half plane
\begin{equation}
	\mathbb{H}_g = \{ \Omega \in M_g(\mathbb{C})~ | ~\Omega^t = \Omega, ~\textrm{Im}(\Omega) > 0\}.
\end{equation}
Equation (\ref{eq:mirrorcurve}) defines a conic fibration over the abelian surface $\mathbb{C}^2/(\mathbb{Z}^2 \oplus \Omega \mathbb{Z}^2)$ with discriminant being the genus $(r+1)N +1$ curve $F^{\textrm{open}} = 0$, and $u$ and $v$ are sections of suitable line bundles over the abelian surface. The curve $F^{\textrm{open}}=0$ is also known as \textit{mirror curve}. It defines a hypersurface in the ambient space $\mathbb{C}^2/(\mathbb{Z}^2 \oplus \Omega \mathbb{Z}^2)$ spanned by $(z_1,z_2)$. From this perspective, the total ambient space of the non-compact Calabi-Yau 3-fold is given by $\mathbb{C}^2/(\mathbb{Z}^2 \oplus \Omega \mathbb{Z}^2) \times \mathbb{C}^* \times \mathbb{C}^*$ with coordinates $(u,v,z_1,z_2)$.
Given the above definitions, our aim in the following sections will be to derive the expression (\ref{eq:mirrorcurve}) from the results obtained in section \ref{sec:ADE}. Note that the main difference to \ref{sec:ADE} is the fact that the base of the elliptic fibration is the blow-up of an affine ADE singularity (in the present case affine A-type) instead of the ordinary one and thus contains an elliptic curve itself. This modification leads to an emerging little string theory in the remaining six dimensions upon compactification of F-theory on the Calabi-Yau $X_{(r+1,N)}$. This has far reaching implications for dualities between quantum field theories in six and five dimensions and we shall devote the next section to their implications for geometry, before turning in the final section to a derivation of (\ref{eq:mirrorcurve}). The broader picture developed in section \ref{sec:littlestring} will then also allow us to study the cases with affine $D$ and $E$ base geometry.

\subsection{Little string theory}
\label{sec:littlestring}

The fiber-base duality encountered in the above discussed geometric picture has a natural interpretation in the context of so called \textit{little string theories} where it also has a generalization to the $D$ and $E$ cases. Let us review this in the following where we shall follow in the first half the presentation of \cite{Ohmori:2015pia} (for related work see \cite{Bhardwaj:2015oru,Kim:2015gha,Aganagic:2015cta,Kim:2017xan}).

\begin{figure}[h!]
  \centering
	\includegraphics[width=0.5\textwidth]{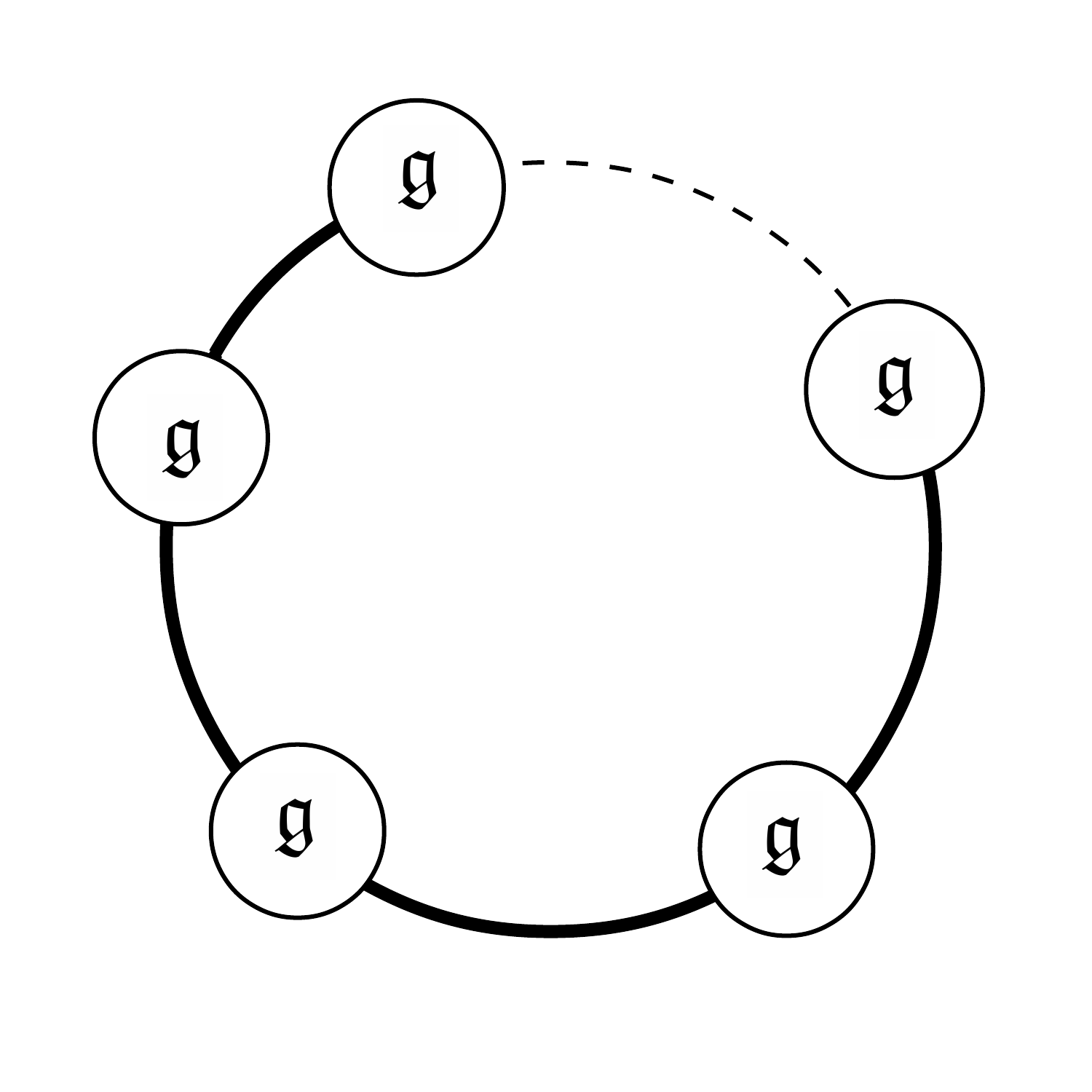}
  \caption{Quiver for the theory $\mathcal{T}_{\frak{g},N}^{\textrm{A}}$ after circle reduction to five dimensions. The thick lines connecting the nodes correspond to generalized bi-fundamental matter.}
  \label{fig:T(A)quiver}
\end{figure}

\begin{figure}[h]
  \centering
  \subfloat[$\mathcal{T}_{A_r,N}^{\textrm{B}}$]{\label{fig:TB1}\includegraphics[width=0.4\textwidth]{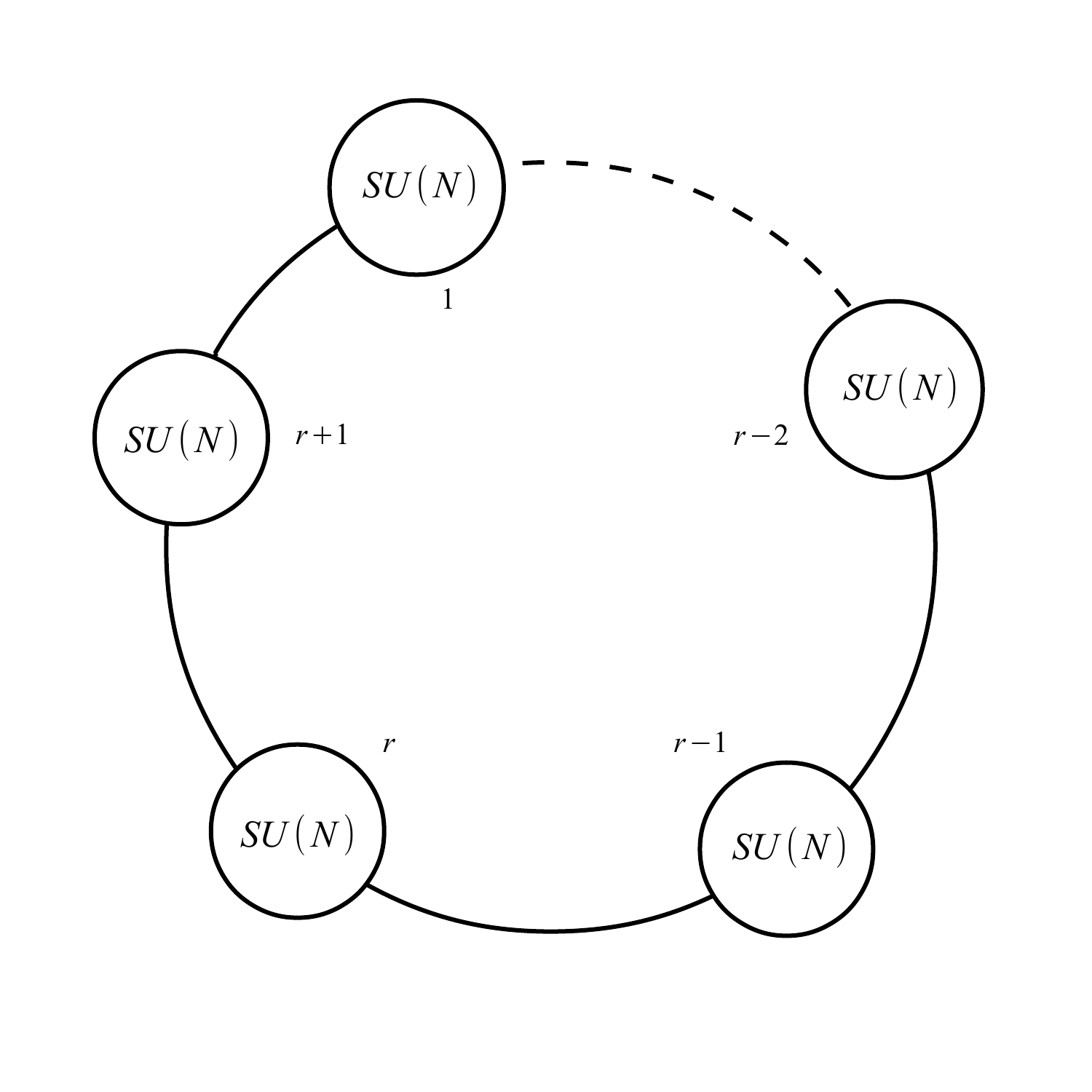}}
  \hspace{.2in}
  \subfloat[$\mathcal{T}^{\textrm{A}}_{A_r,N}$]{\label{fig:TA1}\includegraphics[width=0.4\textwidth]{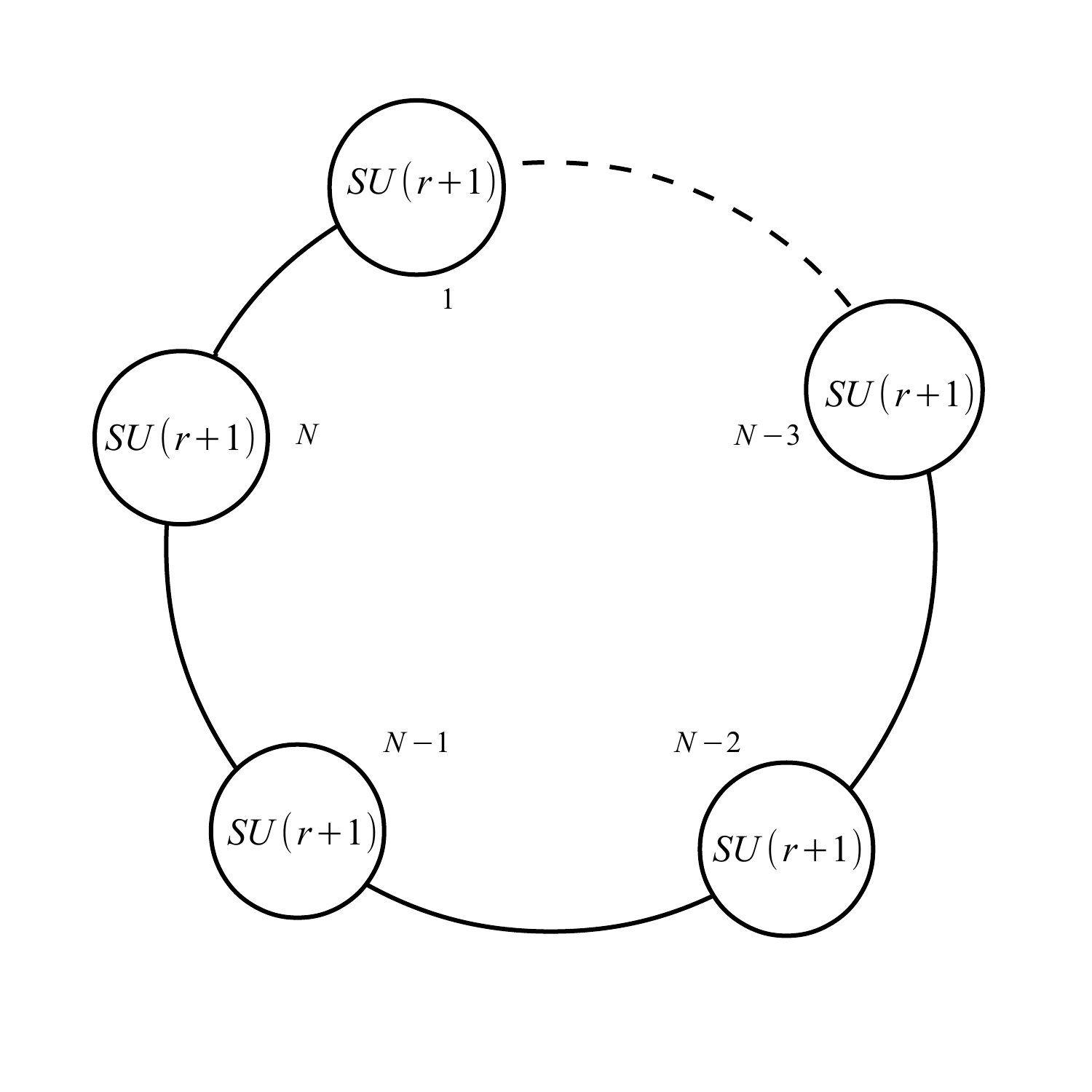}}
  \caption{The fiber-base duality for the case $\frak{g}=A_r$. The nodes in each diagram are connected by free bi-fundamental hypermultiplets.}
  \label{fig:duality1}
\end{figure}

\begin{figure}[h!]
  \centering
	\includegraphics[width=\textwidth]{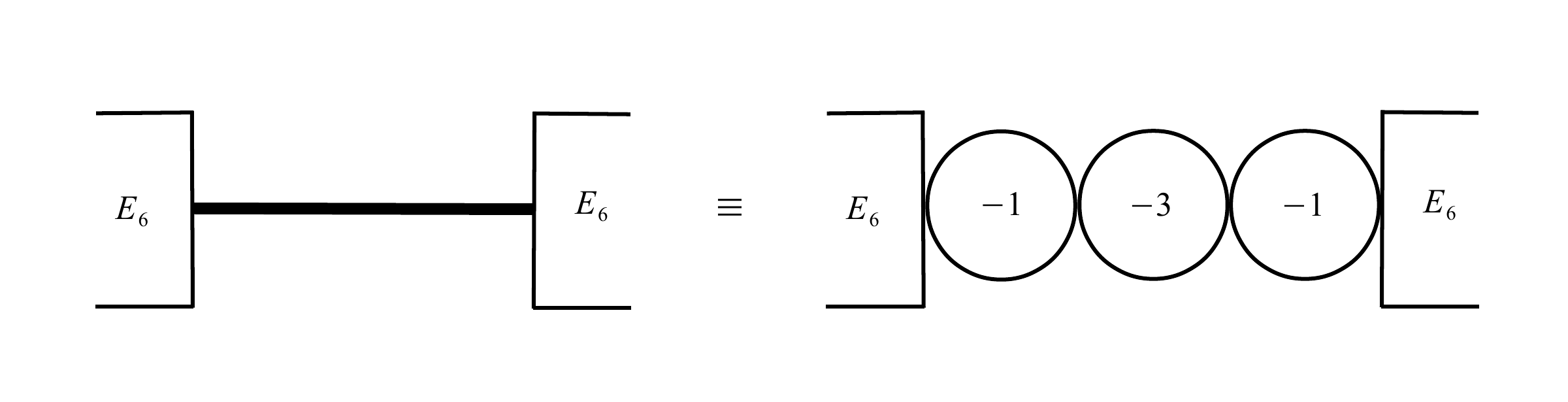}
  \caption{$E_6$ conformal matter.}
  \label{fig:E6conformalmatter}
\end{figure}

\begin{figure}[h]
  \centering
  \subfloat[$\mathcal{T}_{E_6,1}^{\textrm{B}}$]{\label{fig:TB2}\includegraphics[width=0.6\textwidth]{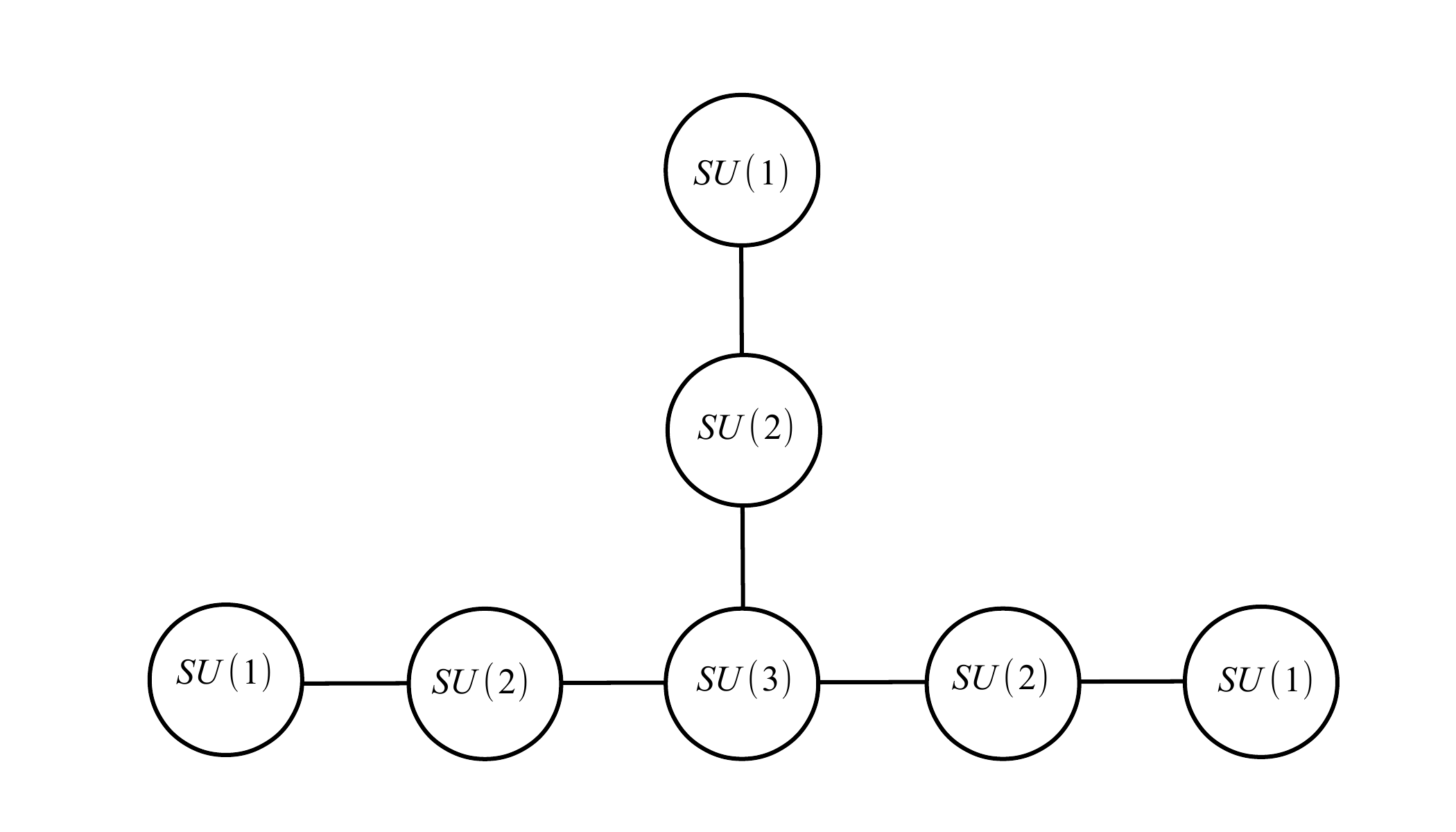}}
  \subfloat[$\mathcal{T}^{\textrm{A}}_{E_6,1}$]{\label{fig:TA2}\includegraphics[width=0.32\textwidth]{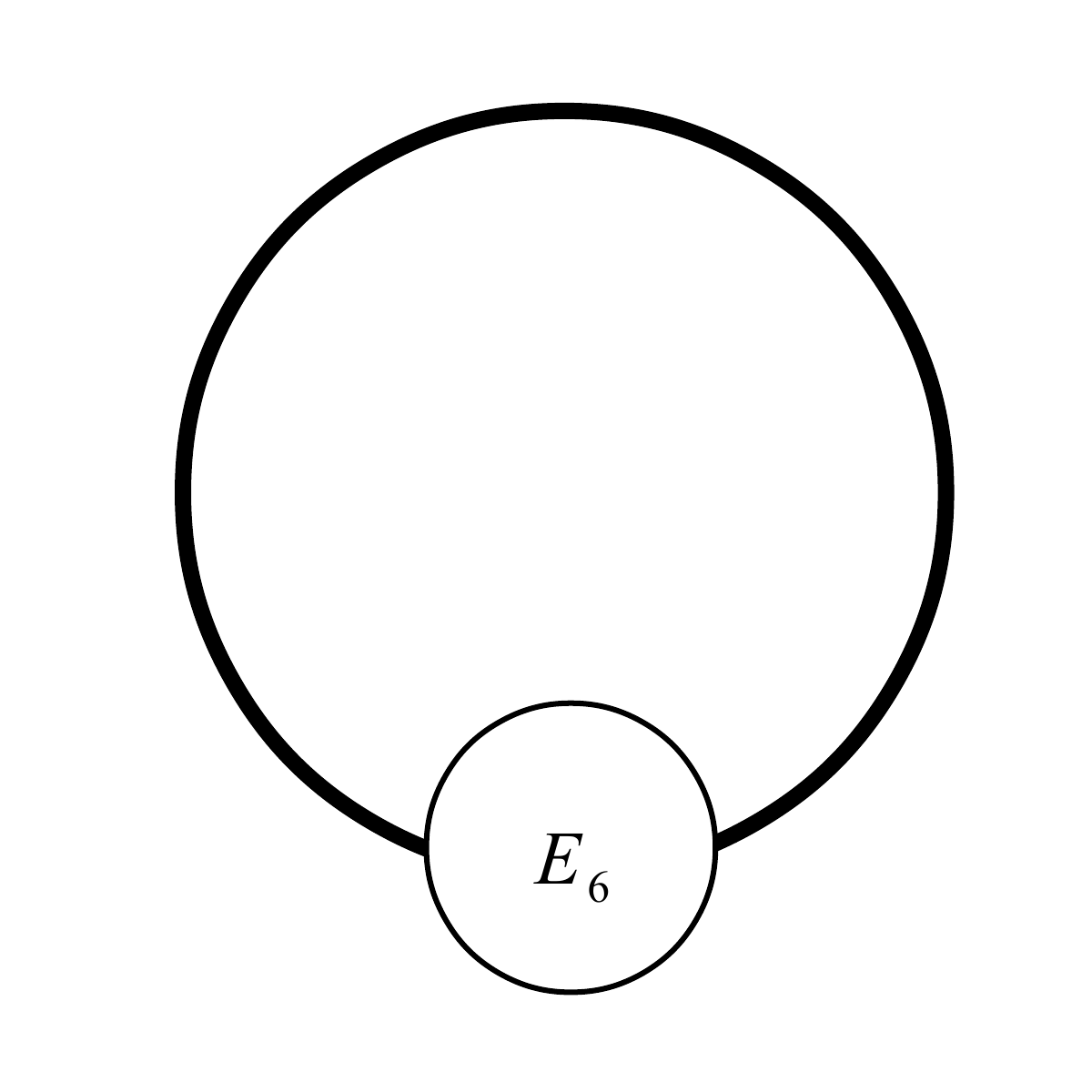}}
  \caption{The fiber-base duality for the case $\frak{g}=E_6$. The nodes in diagram (a) are connected by free bi-fundamental hypermultiplets while the adjoint field of diagram (b) is an $E_6$ conformal matter theory.}
  \label{fig:duality2}
\end{figure}

In the framework of little string theories the fiber-base duality reduces to T-duality of two 6d $\mathcal{N}=(1,0)$ theories. On the one hand, consider type IIB string theory with $N$ NS5-branes on $\mathbb{C}^2/\Gamma_{\frak{g}}$. Here $\frak{g}$ is a general Lie algebra of ADE type. By taking S-duality, this is equivalent to $N$ D5-branes on $\mathbb{C}^2/\Gamma_{\frak{g}}$, and therefore on a generic point on its tensor branch, this theory is the quiver $\prod_{i=0}^{\textrm{rank}\frak{g}} \textrm{SU}(d_i N)$. Equivalently this little string theory is given by our familiar 6d $\mathcal{N}=(1,0)$ SCFT $\mathcal{T}_{\frak{g}}^{\textrm{6d}}\{SU(N)\}$ coupled to an $\textrm{SU}(N)$ vector multiplet. This extra $\textrm{SU}(N)$ vector multiplet then corresponds to gauging the flavor symmetry of the affine node. Let us denote this little string theory by $\mathcal{T}_{\frak{g},N}^{\textrm{B}}$. In the framework of geometric engineering it arises by compactifying F-theory on a Calabi-Yau manifold $X_{N,\frak{g}}$ which admits the following fibration structure
\begin{equation}
	\pi_{1,i} : X_{N,\widehat{\frak{g}}} \rightarrow \widehat{A}_{d_i N-1}, \quad \pi_2 : X_{N,\widehat{\frak{g}}} \rightarrow \widehat{\frak{g}},
\end{equation}
where here $\widehat{\frak{g}}$ denotes the total space of an elliptic fibration over the unit disc $\mathbb{D}$ such that all fibers are smooth except for the central fiber, where the elliptic curve degenerates to a union of nodal curves of the Kodaira type of $\frak{g}$. The fibrations $\pi_{1,i}$ merely state that the $i$'th $\mathbb{P}^1$ in $\widehat{\frak{g}}$ is wrapped by $d_i N$ 7-branes. In this picture, the K\"ahler modulus of each $\mathbb{P}^1$ in $\widehat{\frak{g}}$ is given by $\rho_i$ and the total modulus is given by
\begin{equation}
	\rho = \sum_{i=0}^{\textrm{rank}\frak{g}} d_i \rho_i, \quad \textrm{or equivalently} \quad q_{\rho} = \prod_{i=0}^{\textrm{rank}\frak{g}} q_{\rho_i}^{d_i}.
\end{equation}
Taking the limit $\rho_0 \rightarrow i \infty$ then gives rise to our SCFT $\mathcal{T}_{\frak{g}}^{\textrm{6d}}\{SU(N)\}$ on its tensor branch. In the $A$ case, the $\mathbb{P}^1$ corresponding to the affine node splits into two non-compact curves in this limit which support the flavor symmetries of the resulting SCFT. Now let us turn to the other little string theory $\mathcal{T}_{\frak{g},N}^{\textrm{A}}$ obtained from type IIA string theory with $N$ NS5-branes on $\mathbb{C}^2/\Gamma_{\frak{g}}$. Lifting this to M-theory, we have $N$ M5-branes arranged on points along a circle and probing $\mathbb{C}^2/\Gamma_{\frak{g}}$. In the limit where the radius of the transverse circle is infinite, these theories become the \textit{conformal matter} SCFT's studied in \cite{DelZotto:2014hpa}. In the present case, however, after reduction to five dimensions one obtains a circular quiver with $N$ nodes given by the gauge group of $\frak{g}$ and generalized bifundamental matter studied in \cite{DelZotto:2014hpa} connecting them, as shown in Figure \ref{fig:T(A)quiver}.
The theories $\mathcal{T}_{\frak{g},N}^{\textrm{B}}$ and $\mathcal{T}^{\textrm{A}}_{\frak{g},N}$ are related by T-duality upon compactification on a circle down to five dimensions. In the five-dimensional setting this T-duality then translates to a perfect duality between two seemingly very different quiver gauge theories. One can easily see that the Higgs branches of the two theories match. From the brane constructions presented above, the corresponding Higgs branches are given by moduli spaces of $U(N)$ instantons on the ALE spaces $\mathbb{C}^2/\Gamma_{\frak{g}}$. The matching between the Coulomb branches is, however, more complicated to see and one has to resort to a case by case study. In the following, we shall look at two examples of the duality between these theories. In the first case, we take $\frak{g}$ to be given by $A_r$. Counting parameters, we see that theory $\mathcal{T}_{A_r,N}^{\textrm{B}}$ has $r+1$ coupling constants, $(r+1)(N-1)$ Coulomb parameters, and one mass parameter. Altogether these give $N(r+1)+1$ parameters. On the other hand, theory $\mathcal{T}^{\textrm{A}}_{A_r,N}$ is the circular quiver with $N$ nodes of gauge group $SU(r+1)$. Thus we have $Nr$ Coulomb parameters, $N$ coupling constants and one mass parameter. Again we have $N(r+1)+1$ parameters in total. This situation is depicted in Figure \ref{fig:duality1}.
Note that in both diagrams the bi-fundamental matter fields are ordinary free hypermultiplets. This will be not the case in our second example to which we turn now. Consider the theory $\mathcal{T}^{\textrm{B}}_{E_6,1}$ where for simplicity we are restricting ourselves to the case $N=1$. This theory consists of $SU(d_i)$ gauge nodes for $i=0,\ldots,6$ connected according to the affine $E_6$ Dynkin diagram. Counting parameters, we obtain $7$ coupling constants, and $\sum_i (d_i -1)=5$ Coulomb parameters. The bi-fundamentals have no masses here, as there is no mass-deformation for $D$ and $E$ type quivers. Thus, altogether we arrive at $12$ parameters. The dual $\mathcal{T}^{\textrm{A}}_{E_6,1}$ theory consists of an $E_6$ gauge node and a conformal matter adjoint hypermultiplet. However, as shown in \cite{DelZotto:2014hpa}, $E_6$ conformal matter hypermultiplets are 6d SCFT's themselves and are given in terms of the chain $(-1)(-3)(-1)$ of $-n$ curves. This is graphically depicted in Figure \ref{fig:E6conformalmatter}.
In the dictionary of \cite{Heckman:2013pva} (see also \cite{Haghighat:2014vxa}) the $-1$ curve supports no gauge group whereas the $-3$ curve supports $SU(3)$ gauge symmetry in the bulk of the 6d theory. Therefore, we get $2$ extra Coulomb parameters from the conformal matter in this case. Furthermore, each $\mathbb{P}^1$ contributes a tensor multiplet which in total contribute $4$ more parameters. Combining these with the Coulomb branch parameters of $E_6$ we again obtain $12$ parameters in total. We depict this duality in Figure \ref{fig:duality2}.
The present analysis can be carried over to the remaining Dynkin diagrams and one always finds a match in the count of parameters between the dual theories. This suggests that both theories, namely $\mathcal{T}^{\textrm{A}}$ as well as $\mathcal{T}^{\textrm{B}}$, have the same underlying mirror curve. Now since some coupling parameters of theory $\mathcal{T}^{\textrm{A}}$ appear as Coulomb branch parameters and hence periods of theory $\mathcal{T}^{\textrm{B}}$ and vice verse, one sees that the genus of the mirror curve has to be equal to the total number of parameters in both theories. We shall now turn to the study of this mirror curve.

\subsection{Derivation of mirror curve}
\label{sec:mirrorcurvederivagion}

The goal of this section is to derive expression (\ref{eq:mirrorcurve}) from the thermodynamic limit of the corresponding little string partition function. As explained in the previous section, the Calabi-Yau $X_{(r+1,N)}$ with affine $A$-type elliptic fiber and base corresponds to the little string theory $\mathcal{T}^{\textrm{B}}_{A_r,N}$ which in the limit $\rho \rightarrow i \infty$ gives back our 6d SCFT $\mathcal{T}^{\textrm{6d}}_{A_r}\{\textrm{SU}(N)\}$, where we shall focus on the case without mass-deformation, i.e. $\sigma=0$. From the point of view of theory $A$, in the limit $\rho \rightarrow i \infty$ we obtain the conformal matter theories studies in \cite{DelZotto:2014hpa}. The corresponding Seiberg-Witten theories in this limit were studied in \cite{DelZotto:2015rca} by deforming Landau-Ginzburg orbifold theories. However, the SW curves for Little String theories were not covered in \cite{DelZotto:2015rca} as a corresponding Landau-Ginzburg description is not known. In that sense, the results we will be presenting in this section are more general. It would be interesting to connect our results to the those of \cite{DelZotto:2015rca} which provides algebraic descriptions for the SW curve, whereas we are providing transcendental expressions given by theta functions. We leave this as an interesting open problem for the future.

In the language of quiver gauge theories studied in section \ref{sec:ADE} the difference between the little string theory as compared to the 6d SCFT is that on their tensor branch the little string theory gives rise to an affine $\prod_{i=0}^r \textrm{SU}(N)$ quiver while the SCFT admits an ordinary quiver description $\prod_{i=1}^r \textrm{SU}(N)$. Therefore, in order to study its thermodynamic limit, we have to resort to equation (\ref{eq:affineF0}) such that the partition density $\varrho_0(z)$ is given by equation (\ref{eq:partitiondensity}) for $i=0$. In this theory there will be no masses and it corresponds to the type II classification of quiver gauge theories studied in \cite{Nekrasov:2012xe}. In fact, the authors of \cite{Nekrasov:2012xe} study the geometry emerging in the affine case with the only difference to our present setup being that they use the ordinary multi-gamma functions whereas we have come across their elliptic version in (\ref{eq:Zfinal2}). We will shortly discuss the implications of this which has to do with the fact that our Calabi-Yau $X_{(r+1,N)}$ has not only an elliptic base but also an elliptic fiber. But let us first quote the result of \cite{Nekrasov:2012xe} for the $\widehat{A}_r$ quiver. There, it was shown that the corresponding Seiberg-Witten curve is the zero locus of a section of a degree $r+1$ line bundle over $\mathbb{T}_{\langle w\rangle}$, where
\begin{equation}
	\mathbb{T}_{\langle w \rangle} \equiv \mathbb{C}_{\langle w \rangle}/\{\mathbb{Z} + \rho \mathbb{Z}\}.
\end{equation}
In fact, $\mathbb{T}_{\langle w \rangle}$ is the mirror dual of the elliptic base of the Calabi-Yau $X_{(r+1,N)}$, denoted by $T^2_{\rho}$. The corresponding section is given by\footnote{In this section we are exclusively using the theta function as defined in equation (\ref{eq:theta}).}
\begin{equation}
	s^b(w;\mathbf{w}(z)) = y_0(x) \prod_{i=1}^{r+1} \theta(t/t_i(x);q_{\rho}), \quad t = e^{2\pi i w},~x=e^{2\pi i z}.
\end{equation}
Using identity (7.93) of \cite{Nekrasov:2012xe}, we can write
\begin{equation} \label{eq:sdecomp}
	s^b(w;\mathbf{w}(z)) = \sum_{i=0}^r q_{\rho}^{-\frac{i}{2}} q_{\rho}^{\frac{i^2}{2(r+1)}} \Theta_{i}^{\widehat{A}_r}(y_0(x);\vec{t}(x);q_{\rho}) (-q_{\rho}/t)^i \theta(-(-t)^{r+1} q_{\rho}^{-i};q_{\rho}^{r+1}),
\end{equation}
where we have used that $\Theta_{i}^{\widehat{A}_r}$ are generators of the ring of affine Weyl group invariants defined as
\begin{equation}
	\Theta_{j}^{\widehat{A}_r}(c;\vec{t};q) = c q^{-\frac{j^2}{2(r+1)}} \sum_{\vec{n} \in \Lambda_j} \prod_{k=0}^r t_k^{n_k} q^{\half \sum_{l=0}^r n_l^2}, \quad \Lambda_j = \left\{ n \in \mathbb{Z}^{r+1} |  \sum_{l=0}^r n_l = j\right\}.
\end{equation}
We next define Weyl invariant characters as follows
\begin{equation}
	\widehat{\chi}_i(y_0(x);\vec{t}(x);q_{\rho}) =q_{\rho}^{-\frac{i}{2}} q_{\rho}^{\frac{i^2}{2(r+1)}} \Theta_{i}^{\widehat{A}_r}(y_0(x);\vec{t}(x);q_{\rho}),
\end{equation}
in terms of which the spectral curve equation becomes
\begin{equation} \label{eq:affineArspc}
	\sum_{i=0}^r \widehat{\chi}_i(y_0(x);\vec{t};q_{\rho}) (-q_{\rho}/t)^i \theta(-(-t)^{r+1} q_{\rho}^{-i}; q_{\rho}^{r+1}) = 0.
\end{equation}
Let us now see how we obtain the ordinary $A_r$ spectral curve (\ref{eq:Arspc}) in the limit $q_{\rho} \rightarrow 0$. To this end, observe that
\begin{equation}
	\lim_{q_{\rho} \rightarrow 0} (-q_{\rho}/t)^i \theta(-(-t)^{r+1} q_{\rho}^{-i}; q_{\rho}^{r+1}) = \lim_{q_{\rho} \rightarrow 0} (-q/t)^i (1+(-t)^{r+1} q_{\rho}^{-i}) = (-t)^{r+1-i}.
\end{equation}
Since $i$ is running from $0$ to $r$ we see that we are recovering all powers of $t$ appearing in (\ref{eq:Arspc}) except for the constant term. Here is however a catch, as already mentioned in section \ref{sec:littlestring} the correct limit is taken by splitting the affine node $0$ into two pieces and translated to our current setup this implies that we have to perform the following substitution when taking the non-affine limit:
\begin{equation}
	\widehat{\chi}_0(y_0(x);\vec{t};q_{\rho})\phi_0(t;q_{\rho}) \rightarrow \widehat{\chi}_0(y_0(x); \vec{t};q_{\rho})\phi_0(t;q_{\rho}) + \widehat{\chi}_{r+1}(y_{r+1}(x); \vec{t};q_{\rho}) \phi_{r+1}(t;q_{\rho}),
\end{equation}
where we have defined
\begin{equation}
	\phi_i(t;q) = (-q_{\rho}/t)^i \theta(-(-t)^{r+1} q_{\rho}^{-i}; q_{\rho}^{r+1}).
\end{equation}
Using this substitution, the only remaining task is to show that
\begin{equation}
	\lim_{q_{\rho} \rightarrow 0} \widehat{\chi}_i = \chi_i(y(z)) \zeta(z)^i \prod_{j=1}^{i-1} \mathcal{P}_j^{i-j}(z).
\end{equation}
But this is easily verified by observing the limiting behavior
\begin{equation}
	\lim_{q \rightarrow 0} q^{-\frac{j}{2}} q^{\frac{j^2}{2p}} \Theta^{\widehat{A}_{p-1}}_j(c;\vec{t};q) = c e_j(t_1,\ldots,t_p),
\end{equation}
where the $e_j$ are elementary symmetric polynomials in $p$ variables. We are now ready to apply our master equations (\ref{eq:fibersection}) to the affine Weyl group invariant characters, that is we have
\begin{equation}
	\widehat{\chi}_i(y_0(x);\vec{t}(x);q_{\rho}) = s_i^f(z;\mathbf{z}), \quad z_{i,l} = z_{i,l}(\boldsymbol{\mu},\boldsymbol{\rho}).
\end{equation}
Note that the parameters $\boldsymbol{\mu}$ and $\boldsymbol{\rho}$ now also depend on the affine node $i=0$:
\begin{equation}
	\boldsymbol{\mu} =(\vec{\mu}^{(0)},\ldots, \vec{\mu}^{(r)}), \quad \boldsymbol{\rho} = (\rho_0,\ldots,\rho_r).
\end{equation}
Using
\begin{equation}
	\widehat{\chi}_i = s_i^f = \widehat{T}_{i,0} \prod_{l=1}^N \theta(x/x_{i,l};q_{\tau}), \quad \lim_{q_{\rho} \rightarrow 0} \widehat{T}_{i,0} = T_{i,0},
\end{equation}
we now see that equation (\ref{eq:affineArspc}) becomes
\begin{eqnarray}
	0 & = & \sum_{i=0}^r \widehat{T}_{i,0} \prod_{l=1}^N \theta(x/x_{i,l};q_{\tau})(-q_{\rho}/t)^i \theta(-(-t)^{r+1} q_{\rho}^{-i};q_{\rho}^{r+1}) \nonumber \\
	~ & = & \sum_{i=0}^r  \sum_{l=0}^{N-1} q_{\tau}^{-\frac{l}{2}} q_{\tau}^{\frac{l^2}{2N}} \Theta^{\widehat{A}_{N-1}}_l(\widehat{T}_{i,0};\vec{x}_i;q_{\tau})(-q_{\tau}/x)^l \theta(-(-x)^N q_{\tau}^{-l};q_{\tau}^N) \nonumber \\
	~ & ~ & \times (-q_{\rho}/t)^i \theta(-(-t)^{r+1}q_{\rho}^{-i};q_{\rho}^{r+1}).
\end{eqnarray}
In the second line we have applied equation (\ref{eq:sdecomp}) once again, this time for the section $s^f_i$. On the other hand, the SYZ mirror curve restricted to the case without mass-deformation, namely $\sigma=0$, becomes
\begin{equation}
	0 = \sum_{i=0}^r \sum_{l=0}^{N-1} K_{i,l} \Delta_{i,l} \Theta_2\left[\begin{array}{c}(\frac{i}{r+1},\frac{l}{N})\\(\frac{-(r+1)\rho}{2},-\frac{N\tau}{2})\end{array}\right]\left((r+1)z_1,N z_2; \left[\begin{array}{cc}(r+1)\rho & 0 \\ 0 & N \tau\end{array}\right]\right).
\end{equation}
It is easily derived that
\begin{eqnarray}
	~ & ~ & \Theta_2\left[\begin{array}{c}(\frac{i}{r+1},\frac{l}{N})\\(\frac{-(r+1)\rho}{2},-\frac{N\tau}{2})\end{array}\right]\left((r+1)z_1,N z_2; \left[\begin{array}{cc}(r+1)\rho & 0 \\ 0 & N \tau\end{array}\right]\right) \nonumber \\
	~ & = & \Theta_1\left[\begin{array}{c}\frac{l}{N}\\\frac{-N\tau}{2}\end{array}\right](Nz_2;N\tau)\Theta_1\left[\begin{array}{c}\frac{i}{r+1}\\\frac{-(r+1)\rho}{2}\end{array}\right]((r+1)z_1;(r+1)\rho),
\end{eqnarray}
where $\Theta_1$ is the genus $1$ Riemann theta function. Next, one can show that the following identity holds
\begin{equation}
	\theta(x;q_{\tau}) = -i q_{\tau}^{-\frac{1}{8}} x^{\half} \Theta_1\left[\begin{array}{c}-\half\\-\half\end{array}\right](z;\tau),
\end{equation}
which together with the transformation property
\begin{equation}
	\Theta_1\left[\begin{array}{c}\frac{\epsilon}{2}\\\frac{\epsilon'}{2}\end{array}\right](z+\tau \frac{m}{2}+\frac{n}{2};\tau) = \exp 2\pi i \left(-\half m z - \frac{1}{8} m^2 \tau - \frac{1}{4}m (\epsilon' + n)\right) \Theta_1\left[\begin{array}{c}\frac{\epsilon+m}{2}\\\frac{\epsilon'+n}{2}\end{array}\right](z;\tau),
\end{equation}
gives after a sequence of steps
\begin{equation}
	\theta(-(-x)^N q_{\tau}^{-l};q_{\tau}^N) = q_{\tau}^{-\frac{l}{2}} q_{\tau}^{-\frac{l^2}{2N}}(-x)^l \Theta_1\left[\begin{array}{c}\frac{l}{N}\\ \frac{-N\tau}{2}\end{array}\right](Nz'; N\tau), \quad z' = -z - \half.
\end{equation}
Similarly, we get
\begin{equation}
	\theta(-(-t)^{r+1} q_{\rho}^{-i}; q_{\rho}^{r+1}) = q_{\rho}^{-\frac{i}{2}}q_{\rho}^{-\frac{i^2}{2(r+1)}} (-t)^i \Theta_1\left[\begin{array}{c}\frac{i}{r+1}\\\frac{-(r+1)\rho}{2}\end{array}\right]((r+1)z';(r+1)\rho).
\end{equation}
This completes our proof for the derivation of the mirror curve upon identifying the open Gromov-Witten generating functions
\begin{equation}
	\Delta_{i,l} = \widehat{T}_{i,0} \vartheta^{\widehat{A}_{N-1}}_l(\vec{x}_i;\tau), \quad \vartheta^{\widehat{A}_{N-1}}_l = q_{\tau}^{-\frac{l}{2}} \sum_{\vec{n} \in \Lambda_l} \prod_{k=1}^N x_k^{n_k} q_{\tau}^{\half \vec{n}^2}.
\end{equation}

\subsection{Mirror curves for $D$ and $E$ types}

One can derive mirror curves for $D$ and $E$ types using the same method stated in the previous subsection. The main difference to the $A$ case is that mass deformation is not allowed in the $D$ and $E$ cases, therefore $\sigma$ is always 0 in these cases. Hence the mirror curves of $D$ and $E$ types are hypersurfaces in the direct product $\mathbb{T}_{\rho} \times \mathbb{T}_{\langle z \rangle}$. The section $s^b$, whose zero locus determines the mirror curve, is a section of the determinant bundle $\textrm{det} V$, where $V$ is a vector bundle over $\mathbb{T}_{\langle w \rangle}$ in the fundamental representation of $G_{\mathrm{q}}$.

Mirror curves of $D$ type quivers can be written as
\begin{equation} \label{eq:Dcurve}
0 = \sum_{i}\tilde{T}_i(x;q_\tau)\tilde{M}^{-1}_{ij}(q_\rho)\tilde{\phi}_j(t;q_\rho),
\end{equation}
where
\begin{equation}
\tilde{T}_i(x;q_\tau) = \left(\hat{T}_{i,0} \prod_{l=1}^N \theta(x/x_{i,l}; q_\tau) \right)^2,\phantom{O}i=0,\,1,\,r-1,\,r
\end{equation}
are squares of degree $N$ Jacobi forms on $\mathbb{T}_{\langle z \rangle}$,
and
\begin{equation}
\tilde{T}_i(x;q_\tau) = \hat{T}_{i,0} \prod_{l=1}^{2N} \theta(x/x_{i,l}; q_\tau),\phantom{O}i=2,\, 3,\, \cdots,\,r-2
\end{equation}
are degree $2N$ Jacobi forms on $\mathbb{T}_{\langle z \rangle}$. $\tilde{M}^{-1}$ is the inverse of the mordular transformation matrix,
\begin{equation}
  \begin{split}
    &\tilde{M}_{ij} = \delta_{ij},\,\,i = 2,\ 3,\ \cdots,\ r-2,\\
    &\tilde{M}_{ij} = (-1)^i c^2\sum_{n\in \mathbb{Z}}q^{(2rn+j)^2/4r} + (-1)^i c^2\sum_{n\in \mathbb{Z}}q^{(2rn+j)^2/4r},\,\,i = 0,\ 1,\ j = 1,\,2,\ r-1,\\
    &\tilde{M}_{ij} = (-1)^i c^2\sum_{n\in \mathbb{Z}}q^{(2rn+j)^2/4r},\,\,i = 0,\ 1,\ j = 0,\ r,\\
    &\tilde{M}_{ij} = (-1)^{r-i} c^2\sum_{n\in \mathbb{Z}}q^{(2rn+r+j)^2/4r} + (-1)^i c^2\sum_{n\in \mathbb{Z}}q^{(2rn+r+j)^2/4r},\,\,i = r-1,\ r,\ j = 1,\,2,\ r-1,\\
    &\tilde{M}_{ij} = (-1)^{r-i} c^2\sum_{n\in \mathbb{Z}}q^{(2rn+r+j)^2/4r},\,\,i = r-1,\ r,\ j = 0,\ r,
  \end{split}
\end{equation}
Finally $\tilde{\phi}_j(t;q_\rho)$ are defined as
\begin{equation}
\begin{split}
&\tilde{\phi}_0(t;q_\rho) = \frac{\theta(-t^{2r}q_\rho;q^{2r})}{\theta(t;q_\rho)^{2r}},\\
&\tilde{\phi}_i(t;q_\rho) = (-1)^i\left(q^{\frac{i}{2}+\frac{i^2}{4r}}_\rho t^{-i}\frac{\theta(-t^{2r}q^{-i}_\rho;q^{2r}_\rho)}{\theta(t;q_\rho)^{2r}} +
q^{2r-\frac{3}{2}i +\frac{i^2}{4r}}_\rho t^{-2r + i}\frac{\theta(-t^{2r}q^{-2r + i}_\rho;q^{2r}_\rho)}{\theta(t;q_\rho)^{2r}}  \right),\\
&\tilde{\phi}_{r}(t;q_\rho) = (-1)^r q^{\frac{3}{4}r}_\rho t^{-r} \frac{\theta(-t^{2r}q^{-r}_\rho;q^{2r}_\rho)}{\theta(t;q_\rho)^{2r}}.
\end{split}
\end{equation}
By construction, the mirror curve (\ref{eq:Dcurve}) has genus $g_D = 2N (r-2) + 4N$.

The mirror curves of E type quivers can be obtained in a similar way. They are defined by the following sets of equations,
\begin{equation} \label{eq:Ecurve}
\begin{split}
&-Y^2+4X^3-g_2(q_\rho)X-g_3(q_\rho)=0,\\
&C^{E_i}(X,Y;g_2(q_\rho), g_3(q_\rho); p_0,\cdots,p_i) = 0, \, i = 6, \,7,\,8.
\end{split}
\end{equation}
$g_2(q_\rho)$ and $g_3(q_\rho)$ are Weierstrass parameters,
\begin{equation}
\begin{split}
&g_2(q)=\frac{1}{12} + 20q+180q^2+560q^3+1460q^4+\cdots,\\
&g_3(q)=-\frac{1}{216} + \frac{7}{3}q + 77 q^2 + \frac{1708}{3}q^3 + \frac{7399}{3}q^4 + \cdots.
\end{split}
\end{equation}
$C^{E_r}$'s are polynomials in $X$ and $Y$ with polynomial coefficients in $g_2$, $g_3$, $p_0$, ...$p_r$ with $r$ being 6, 7, 8. The explicit form can be found in appendix E of  \cite{Nekrasov:2012xe}. Since both base and fiber are elliptic in our case, $p_j$'s are related by modular transformation matrices (see \cite{Nekrasov:2012xe} for details) to sections of line bundles of degree $N d_j$ over $\mathbb{T}_{\langle z \rangle}$,
\begin{equation}
s^f_i= \widehat{T}_{i,0} \prod_{l = 1}^{N d_i} \theta(x/x_{i,l};q_\tau), \,\,i = 0,\,1,\,\cdots, \, r.
\end{equation}
By construction, the mirror curve (\ref{eq:Ecurve}) has genus $g_{E_a} = \sum_i N d_i = N h_a$ for $a=6,7,8$, where $h_a$ is the dual Coxeter number of $E_a$ and given by $h_6 = 12$, $h_7=18$ and $h_8 = 30$.

\section*{Acknowledgment}

We would like to thank A. Kanazawa, C. Kozcaz, S.-C. Lau, E. Looijenga, V. Pestun and S. Vandoren for valuable discussions. BH would like to thank the Institute for Theoretical Physics at Utrecht University for hospitality where part of this work was completed. The work of BH is supported by Yau Mathematical Sciences Center, Tsinghua University. The work of WY is supported by Yau Mathematical Sciences Center, Tsinghua University and the Center for Mathematical Sciences and Applications at Harvard University.

\appendix

\section{Elliptic multi-gamma functions}

In this section we collect basic definitions and properties of multiple elliptic gamma functions following the exposition of \cite{Narukawa}.

Let $x = e^{2\pi i z}$, $q_j = e^{2\pi i \tau_j}$ for $z \in \mathbb{C}$ and $\tau_j \in \mathbb{C} - \mathbb{R}$, and
\begin{equation}
	\underline{q} = (q_0, \cdots, q_r).
\end{equation}
Next, for $\textrm{Im}\tau_j > 0$ for all $j$, define
\begin{equation}
	(x;\underline{q})_{\infty}^{(r)} = \prod_{j_0,\cdots,j_r=0}^{\infty} (1 - x q_0^{j_0} \cdots q_r^{j_r}).
\end{equation}
This infinite product converges absolutely when $|q_j| <1$. It can be shown (see \cite{Narukawa} for more details) that the definition of $(x;q)_{\infty}^{(r)}$ can be analytically continued to other values of $\tau$. Also, note that the function is invariant under an arbitrary permutation of $q_0, \ldots, q_r$.  We next denote
\begin{eqnarray}
	~ & ~ & \underline{\tau} = (\tau_0,\ldots, \tau_r), \nonumber \\
	~ & ~ & \underline{\tau}^-(j) = (\tau_0,\ldots, \check{\tau}_j,\ldots, \tau_r), \nonumber \\
	~ & ~ & \underline{\tau}[j] = (\tau_0,\ldots,-\tau_j,\ldots,\tau_r), \nonumber \\
	~ & ~ & -\underline{\tau} = (-\tau_0,\ldots,-\tau_r), \nonumber \\
	~ & ~ & |\underline{\tau}| = \tau_0 + \cdots + \tau_r, \nonumber
\end{eqnarray}
where by $\check{\tau}_j$ we mean that the entry $\tau_j$ has been omitted. Now we are in the position to define the multiple elliptic gamma function
\begin{equation}
	G_r(z |\underline{\tau}) = (x^{-1} q_0 \cdots q_r; \underline{q})_{\infty}^{(r)} \{(x;\underline{q})_{\infty}^{(r)}\}^{(-1)^r}.
\end{equation}
The hierarchy of $G_r(z|\underline{\tau})$ includes the theta function $\theta(z,\tau)$ and the elliptic gamma function $\Gamma(z,\tau,\sigma)$ which for $\textrm{Im}\tau, \textrm{Im}\sigma >0$ are defined as
\begin{eqnarray}
 \theta(z,\tau) & = & \prod_{j=0}^{\infty} (1-e^{2\pi i((j+1)\tau-z)})(1-e^{2\pi i(j\tau+z)}) = G_0(z|\tau), \nonumber \\
 \Gamma(z,\tau,\sigma) & = & \prod_{j,k=0}^{\infty} \frac{1-e^{2\pi i((j+1)\tau+(k+1)\sigma-z)}}{1-e^{2\pi i(j\tau +k\sigma+z)}} = G_1(z|\tau,\sigma).
\end{eqnarray}
Furthermore, from the definition of $(x;\underline{q})_{\infty}^{(r)}$ one can deduce the following functional equations:
\begin{eqnarray}
	G_r(z+1|\underline{\tau}) & = & G_r(z|\underline{\tau}), \nonumber \\
	G_r(z+\tau_j|\underline{\tau}) & = & G_{r-1}(z|\underline{\tau}^-(j))G_r(z|\underline{\tau}), \nonumber \\
	G_r(-z|-\underline{\tau}) & = & \frac{1}{G_r(z|\underline{\tau})}, \nonumber \\
	G_r(z|\underline{\tau}) & = & \frac{1}{G_r(z-\tau_j|\underline{\tau}[j])}, \nonumber \\
	G_r(z|\underline{\tau})G_r(z|\underline{\tau}[j]) & = & \frac{1}{G_{r-1}(z|\underline{\tau}^-(j))}.
\end{eqnarray}
Using the second equation above, we can compute for $G_2(z|\tau,\epsilon_1,\epsilon_2)$:
\begin{eqnarray}
	~ & ~ & \frac{G_2(z+\epsilon_1 + \epsilon_2|\tau,\epsilon_1,\epsilon_2)}{G_2(z+\epsilon_2|\tau,\epsilon_1,\epsilon_2)}\left/\frac{G_2(z+\epsilon_1|\tau,\epsilon_1,\epsilon_2)}{G_2(z|\tau,\epsilon_1,\epsilon_2)}\right. \nonumber \\
	~ & = & G_1(z+\epsilon_2|\tau,\epsilon_2)/G_1(z|\tau,\epsilon_2) \nonumber \\
	~ & = & G_0(z|\tau). \label{eq:Grec}
\end{eqnarray}
Thus, defining
\begin{equation}
	\gamma(z;\epsilon_1,\epsilon_2) := \log(G_2(z|\tau,\epsilon_1,\epsilon_2),
\end{equation}
(\ref{eq:Grec}) becomes equivalent to the following difference equation
\begin{equation} \label{eq:refinedrecursion}
	\gamma(z+\epsilon_1 + \epsilon_2;\epsilon_1,\epsilon_2) + \gamma(z;\epsilon_1,\epsilon_2)-\gamma(z+\epsilon_1;\epsilon_1,\epsilon_2) - \gamma(z+\epsilon_2;\epsilon_1,\epsilon_2) =\log{\theta(z,\tau)}, 
\end{equation}
which in the unrefined limit $\epsilon_1 = - \epsilon_2 = \hbar$ with $\gamma(z;\hbar) := \gamma(z;\hbar,-\hbar)$ becomes equation (\ref{eq:identity2})\footnote{Note that this gives equation (\ref{eq:identity2}) with the theta function as defined in (\ref{eq:theta}). In order to obtain the logarithm of the theta function (\ref{eq:originaltheta}), one has to perform the substitution $\gamma(z;\hbar) \rightarrow \gamma(z;\hbar) - \frac{x^2}{12 \hbar^2} + \frac{x}{24}$, where $x = e^{2\pi i z}$.}.

\end{document}